\documentclass[12pt]{article}
\usepackage{amssymb}

\newcommand{\be}{\begin{equation}}
\newcommand{\ee}{\end{equation}}
\def\bea{\begin{eqnarray}}
\def\eea{\end{eqnarray}}

\def \del{\partial}

\newcommand{\bn}{\begin{eqnarray}}
\newcommand{\en}{\end{eqnarray}}

\newcommand{\odl}{{\overline{\delta}}}

\newcommand{\p}{\partial}

\newcommand{\nn}{\nonumber}

\newcommand{\no}{\noindent}

\newcommand{\orho}{\overline{\rho}}
\newcommand{\oGamma}{\overline{\Gamma}}
\newcommand{\ogamma}{\overline{\gamma}}
\newcommand{\ophi}{\overline{\phi}}
\newcommand{\oL}{\overline{\Lambda}}
\newcommand{\oi}{\overline{i}}
\newcommand{\oI}{\overline{I}}
\newcommand{\tp}{I_{2s}}
\newcommand{\tpi}{i_{2s}}
\newcommand{\md}{I_{2s-1}}
\newcommand{\mdi}{i_{2s-1}}
\newcommand{\mdb}{\overline{I}_{2s-1}}
\newcommand{\mdbi}{\overline{i}_{2s-1}}
\newcommand{\bt}{\overline{I}_{2s-2}}

\newcommand{\lsdt}{{\cal L}_{2s}^{SD}}
\newcommand{\lst}{{\cal L}_{2s}^{(s)}}

\newcommand{\lsdm}{{\cal L}_{2s-1}^{SD}}
\newcommand{\lsm}{{\cal L}_{2s-1}^{(s)}}
\newcommand{\lsb}{{\cal L}_{2s-2}^{(s)}}
\newcommand{\ldt}{{\cal L}_{2s}^{D}}

\newcommand{\hth}{\hat{\theta}}

\newcommand{\hp}{\hat{\partial}}
\newcommand{\ns}{\nabla^2}

\def\bea{\begin{eqnarray}}
\def\eea{\end{eqnarray}}

\newcommand{\beq}{\begin{eqnarray}}
\newcommand{\eeq}{\end{eqnarray}}

\begin{document}

\title{\textbf{On higher spin analogues of linearized Topologically Massive Gravity and
linearized ``New Massive Gravity''}}
\author{D. Dalmazi$^{1}$\footnote{dalmazi@feg.unesp.br}, A. L. R. dos Santos$^{2}$\footnote{alessandroribeiros@yahoo.com.br} \\
\textit{{1- UNESP - Campus de Guaratinguet\'a - DFI} }\\
\textit{{CEP 12516-410 - Guaratinguet\'a - SP - Brazil.} }\\
\textit{{2- Instituto Tecnol\'ogico de Aeron\'autica, DCTA} }\\
\textit{{CEP 12228-900, S\~ao Jos\'e dos Campos - SP - Brazil} }\\}
\date{\today}
\maketitle

\begin{abstract}

We suggest a new spin-4 self-dual model  (parity singlet) and a new spin-4  parity doublet in $D=2+1$. They are of higher order in derivatives and are described by a totally symmetric rank-4 tensor without extra auxiliary fields. Despite the higher derivatives they are ghost free.  We find gauge invariant field combinations which allow us to show that the canonical structure of the spin-4 (spin-3) models follows the same pattern of its spin-2 (spin-1) counterpart after field redefinitions. For $s=1,2,3,4$, the spin-$s$ self-dual models of order $2s-1$  and the doublet models of order $2s$ can be written in terms of three gauge invariants. The cases $s=3$ and $s=4$ suggest  a restricted conformal higher spin symmetry as a principle for defining linearized topologically massive gravity and linearized ``New Massive Gravity'' for arbitrary integer spins. A key role in our approach is played by the fact that the Cotton tensor in $D=2+1$ has only two independent components for any integer spin.


\end{abstract}

\newpage

\section{ Introduction}

Contrary to the real world in $D=3+1$ where local actions for massless particles of spin-s necessarily describe both helicties $\pm s$, in $D=2+1$  there are  local actions for each helicity $+s$ or $-s$, they may be called self-dual models or parity singlets and represent now massive particles. The Maxwell Chern-Simons (MCS) theory and the linearized topologically massive gravity (TMG) \cite{djt} are paradigmatic examples of self-dual models of spin-1 and spin-2 respectively. By means of a soldering procedure \cite{stone}, see also \cite{review}, it is possible to join together opposite helicities into a parity invariant (parity doublet) local action with helicities $+s$ and $-s$. In the spin-1 case we obtain the  Maxwell-Proca theory \cite{bk,gs} while the soldering   of spin-2 second order (in derivatives) self-dual models \cite{dmc} lead to the massive spin-2 Fierz-Pauli theory \cite{fp}, see \cite{dm1}. Since those massive actions  have the same form in arbitrary dimensions we may say that the self-dual models in $D=2+1$ work like building blocks of those massive particles in arbitrary $D$ dimensions.

Another connection with higher dimensions comes from the fact that massive models may be deduced via Kaluza-Klein dimensional reduction of massless particles, see \cite{rs}. In particular, the self-dual models in $D=2+1$ may be obtained from massless particles in $D=3+1$ as shown in \cite{sdr} for $s=1,2,3$. Such procedure leads to first order self-dual models   which require in general auxiliary fields in order to produce the so called Fierz-Pauli conditions. The auxiliary fields may turn into dynamical fields and become obstacles when interactions are considered. It is possible however, to trade auxiliary fields in higher derivatives and gauge symmetries necessary to eliminate ghosts. Those symmetries may be used as a guiding principle for the introduction of interactions. Here we are specially interested in those higher derivative gauge invariant higher spin models.

For each spin-s there seem to be a ``$2s$ rule'' in $D=2+1$  such that we have ghost free self-dual models of $j$-th order in derivatives with $j=1,2, \cdots, 2s$. By means of a Noether gauge embedding (NGE) procedure \cite{anacleto} we can systematically climb up from the $j$-th to the $(j+1)$-th order from bottom $(j=1)$ to top $(j=2s)$, stepwise eliminating auxiliary fields and adding gauge symmetries.  The procedure works nicely for $s=1,3/2,2$, see \cite{gs,mls,sd4} respectively, but at $s=3$ it is  only partially successful. In \cite{nge3} we go from $j=1$ until $j=4$, but we have not been able to connect the spin-3 fourth-order model of \cite{nge3}, containing auxiliary fields, with the top 6th order self-dual model of  \cite{bht_hd} which has no auxiliary fields.

Since the top model of order $2s$ is known for arbitrary integer \cite{yin} and half-integer \cite{kuzenko18} spin-s we might try as an alternative approach to climb down the ladder of derivatives. This is what we pursue in the present work. We are able to go one step down from the top for spins $s=3$ and $s=4$ without introducing auxiliary fields. We believe that our approach can be generalized for arbitrary integer spins. In the spin-4 case we obtain a new 7th ($2s-1$) order self-dual model and a new 8th ($2s$) order doublet model. We show in a gauge invariant way that they are both ghost free. They correspond respectively to the spin-4 analogues of the spin-2 linearized TMG\footnote{The authors of \cite{kuzenko18} have also suggested a higher spin ``topologically massive'' theory of order $2s-1$ in $D=2+1$ but it requires furthter auxiliary fields, differently from ours.}   and of the linearized ``New Massive Gravity'' (NMG) of \cite{bht} respectively.

Although not explicitly Lorentz invariant, we employ here a formalism based on gauge invariants which dispenses the use of gauge conditions . The absence of gauge conditions allows us to show that the canonical structure of the spin-4 (spin-3) case basically coincides with the canonical structure of the lower spin-2 (spin-1) case. Our approach might be useful in investigating other higher derivative models.

\section{General set up}

Throughout this work\footnote{ We only work on the flat space and use $\eta_{\mu\nu} = (-,+,+)$. Symmetrizations do not contain numerical factors, e.g., $(\alpha\beta)=\alpha\beta + \beta\alpha $ and $(\alpha\beta\gamma)=\alpha\beta\gamma + \beta\gamma\alpha + \gamma\alpha\beta  $.} we will be using three Lagrangians ${\cal L}_{k}^{(s)}$ of k-th order in derivatives with $k=2s,2s-1,2s-2$ as basic ingredients for building up spin-s self-dual  models\footnote{In all sections the lower index in the Lagrangian symbol stands for its order in derivatives, i.e., the highest number of space time  derivatives of the rank-s fundamental field $h_{\mu_1 \cdots \mu_s}$.}  $\lsdt,\lsdm$ and  the doublet model $\ldt$,

\bea S_{2s} &=& b_0 \int \, d^3\, x  \, h_{\mu_1  \cdots \mu_s} E^{\mu_1}_{\!\quad\rho}\, C^{\rho\mu_2\cdots \mu_s } \equiv \int \, d^3\, x \, \lst \quad , \label{lst} \\
S_{2s-1} &=& c_0 \int \, d^3\, x  \,  h_{\mu_1  \cdots \mu_s} \, C^{\mu_1\cdots \mu_s } \equiv \int \, d^3\, x \, \lsm \quad ,\label{lsm} \\  S_{2s-2} &=& d_0 \int \, d^3\, x \, h_{\mu_1  \cdots \mu_s} \, D^{\mu_1\cdots \mu_s } \equiv \int \, d^3\, x \,  \lsb \quad , \label{lsb} \eea

\no where $(b_0,c_0,d_0)$ are arbitrary overall constants and $h_{\mu_1\cdots \mu_s}$ is our fundamental rank-s field, traceful and symmetric $h_{\mu_1\cdots \mu_s}=h_{(\mu_1\cdots \mu_s )}$.  We frequently use

\be E^{\rho\delta} \equiv \epsilon^{\rho\delta\sigma}\p_{\sigma} \quad ; \quad  \Box \theta_{\rho\sigma} \equiv
\Box \,\eta_{\rho\sigma} - \p_{\rho} \p_{\sigma}\quad ; \quad  E^{\mu\nu} E ^{\alpha\beta} = \Box \left( \theta^{\mu\beta}\theta^{\nu\alpha} -
\theta^{\mu\alpha}\theta^{\nu\beta} \right) \quad . \label{id1} \ee

\no  A major role is played by the spin-s Cotton tensor $C_{\mu_1 \cdots \mu_s}$. More specifically in $D=2+1$, it appears in \cite{djt} and \cite{damour} in the spin-2 and spin-3 cases respectively, and for arbitrary integer spin in \cite{pt}. It is of order $2s-1$ in derivatives ($C\sim \p^{2s-1}h$), fully symmetric, transverse and traceless,

\be C_{\mu_1 \cdots \mu_s} = C_{(\mu_1 \cdots \mu_s)} \quad ; \quad \p^{\rho}C_{\rho\mu_2 \cdots \mu_s} = 0 \quad ; \quad \eta^{\rho\nu}C_{\rho\nu\mu_3 \cdots \mu_s} = 0 \label{fpc}\ee

\no Later on we give an explicit formula for $C_{\mu_1 \cdots \mu_s}$ in the flat space. An extension for the $AdS_3$ space including half-integer spins is given in \cite{kuzenko21}. The tensor $D_{\mu_1\cdots \mu_s }$ is of order $2s-2$ in derivatives, fully symmetric too. It is connected with the Cotton tensor via a symmetrized curl,

\be C_{\mu_1 \cdots \mu_s} = E_{(\mu_1}^{\!\quad\rho}D_{\rho\mu_2\cdots \mu_s )} \quad. \label{dtensor} \ee

\no In general there is a multi parametric family of D-tensors satisfying (\ref{dtensor}). We are specially interested in the subset of Lagrangians $\lsb$ without particle content.

We first recall the construction  of the highest order self-dual model $\lsdt $, see \cite{bht_hd} for spin-3, \cite{yin} for arbitrary integer spin and \cite{kuzenko18} for arbitrary half-integer. For arbitrary integer spin-s it is given by a linear combination of $\lst$ and $\lsm$. If we choose $c_0 = - m\ b_0$ we have,

\be \lsdt = b_0 \left\lbrack  h_{\mu_1  \cdots \mu_s} E^{\mu_1}_{\!\quad\rho}\, C^{\rho\mu_2\cdots \mu_s } - m\, h_{\mu_1  \cdots \mu_s} \, C^{\mu_1\cdots \mu_s } \right\rbrack \quad , \label{lsdt} \ee

The corresponding equations of motion,

\be E^{(\mu_1}_{\!\quad\rho}\, C^{\rho\mu_2\cdots \mu_s ) }= m\, s \, C^{\mu_1\cdots \mu_s } \quad , \label{pl} \ee

\no play the role of the Pauli-Lubanski eigenvalue equation in $D=2+1$. If we apply $E^{\gamma}_{\!\quad\mu_1}$ on (\ref{pl}) and use (\ref{id1}), (\ref{fpc}) and (\ref{pl}) recursively, we deduce the Klein-Gordon equations:

\be (\Box - m^2)C_{\mu_1  \cdots \mu_s} = 0 \quad . \label{kg1} \ee

\no It can be shown from first principles that the Fierz-Pauli constraints (\ref{fpc}) and the dynamic equations (\ref{pl}) and (\ref{kg1}) is all we need to have massive particles with helicity $ s \vert m\vert/m$. However, since we have in general higher order time derivatives there might be further particles, including ghosts, so the particle content of (\ref{lsdt}) must be thoroughly investigated. The Lagrangians $\lst$, $\lsm$ and consequently $\lsdt$ are invariant under a  large set of local transformations:

\be \delta h_{\mu_1  \cdots \mu_s} = \p_{(\mu_1}\Lambda_{ \mu_2  \cdots \mu_s)} + \eta_{(\mu_1\mu_2} \psi_{\mu_3  \cdots \mu_s )} \quad ,\label{gt1}\ee

\no where the gauge parameters $\Lambda_{ \mu_1  \cdots \mu_{s-1}}$ and $\psi_{ \mu_1  \cdots \mu_{s-2}}$ are fully symmetric but otherwise arbitrary tensors. Because of those symmetries one can fix convenient gauges and prove that $\lsdt$ only contains massive particles of helicity $+s$ or $-s$ depending on the sign of  $m$, see \cite{bht_hd} and \cite{yin} for the spin-3 and spin-4 cases respectively. The approach we use here allows us to prove the absence of ghosts in the spin $s=3,4$ cases in an off-shell and gauge invariant way as we will see later.

Inspired on $\lsdt$ we define the lower order self-dual model $\lsdm$ combining $\lsm$ and $\lsb$ with $d_0=-s\,m\, c_0$,


\be \lsdm = c _0 \left\lbrack  h_{\mu_1  \cdots \mu_s} E^{(\mu_1}_{\!\quad\rho}\, D^{\rho\mu_2\cdots \mu_s )} - m\, s\, h_{\mu_1  \cdots \mu_s} \, D^{\mu_1\cdots \mu_s } \right\rbrack \quad , \label{lsdm} \ee

\no The equations of motion are given by

\be E^{(\mu_1}_{\!\quad\rho}\, D^{\rho\mu_2\cdots \mu_s ) }= m\, s \, D^{\mu_1\cdots \mu_s } \quad , \label{pl2} \ee

\no Since the left hand side of (\ref{pl2}) is the Cotton tensor, by taking the trace and applying a derivative on (\ref{pl2}) we deduce, with help of the identities (\ref{fpc}), the Fierz-Pauli constraints $\p^{\mu_1}D_{\mu_1\cdots \mu_s }=0$ and
$\eta^{\mu_1\mu_2}D_{\mu_1\mu_2 \cdots \mu_s}=0$ which are now dynamic equations instead of trivial identities as  (\ref{fpc}).

 The application of the curl $E^{\gamma}_{\!\!\!\quad\mu_1}$ on (\ref{pl2}) will similarly lead to (\ref{kg1}) with $C_{\mu_1\mu_2 \cdots \mu_s}$ replaced by $D_{\mu_1\mu_2 \cdots \mu_s}$ which confirms that $\lsdm$ contains massive particles of helicity $s \vert m\vert /m $. There is however, no guarantee that no other propagating particles are present. In the four cases $s=1,3/2,2,3$, see \cite{djt}, \cite{helder}, \cite{sd4} and \cite{sd5} respectively, there is a master action connecting $\lsdm$ with $\lsdt$ with a $D$-tensor satisfying (\ref{dtensor}) and such that $\lsb$ has no particle content. For instance, in the  $s=2$ case there are two choices for the D-tensor, one corresponds to the linearized Einstein-Hilbert theory and the other one to the WTDiff model or linearized unimodular gravity, both Lagrangians have no propagating modes in $D=2+1$. The respective self-dual models $\lsdm$ are the linearized topologically massive gravity \cite{djt} and linearized unimodular topologically massive gravity \cite{ghosh}. One can also combine $\lst$ and $\lsb$ and build up doublet models $\ldt$ containing both helicities $+s$ {\bf and} $-s$. They represent higher spin analogues of the linearized NMG \cite{bht} and of the linearized unimodular NMG \cite{ghosh}. In the next section,
as a preparation for the section 4 where possible choices for $D_{\mu_1\mu_2 \cdots \mu_s}$ will be discussed, we give closed formulae for the Cotton tensor and its symmetrized curl on the flat space. They are convenient for our approach based on the use of gauge invariant field combinations.

\section{The Cotton tensor and the Lagrangians $\lsm$ and $\lst$ }

One can think of  $\lsm$ as the most general spin-s parity odd and Lorentz invariant expression of order $2s-1$ in derivatives invariant under (\ref{gt1}). We start with an Ansatz such that invariance under the higher spin analogue of linearized Diffeomorphisms (Diff)
$\delta h_{\mu_1\cdots \mu_s} = \p_{(\mu_1}\Lambda_{\mu_2\cdots \mu_s)}$
is granted, namely,

\bea \lsm & =& h_{\mu_1\cdots \mu_s}\Box^{s-1} E^{\mu_1\nu_1} \left\lbrack c_0 \,
\theta^{\mu_2\nu_2} \cdots \theta^{\mu_s\nu_s} +  c_1 \,
\theta^{\mu_2\mu_3}\theta^{\nu_2\nu_3}\theta^{\mu_4\nu_4} \cdots \theta^{\mu_s\nu_s} + \cdots \cdots \cdots \right\rbrack h_{\nu_1\cdots \nu_s}  \nn\\
&=& h_{\mu_1\cdots \mu_s}\Box^{s-1} E^{\mu_1\nu_1} \left\lbrack c_0 \theta^{s-1} + c_1 \hth^2 \theta^{s-3} + c_2 \hth^4 \theta^{s-5}+ \cdots\cdots\cdots \right\rbrack^{\mu_2\cdots \mu_s\, \nu_2\cdots \nu_s} h_{\nu_1\cdots \nu_s} , \label{l2s-1}\eea

\no Where $c_j$ with $ j=0,1, \cdots, \left\lbrack \frac{s-1}2\right\rbrack$ are to be determined and
 $\hth$ stands for the transverse operator $\theta_{\mu_j\mu_{j+1}}$ or $\theta_{\nu_j\nu_{j+1}}$ whose indices are contracted within indices of the same h-field. Under generalized Weyl transformations $\delta h_{\nu_1\cdots \nu_s} = \eta_{(\nu_1\nu_2}\psi_{\nu_3\cdots \nu_{s})}$ we have the following structure (suppressing indices)

\be \delta\lsm =   h\, \Box^{s-1} E \left( C_1 \hth \theta^{s-3}\psi  + C_2 \hth^3 \theta^{s-5}\psi  + \cdots\cdots\cdots \right), \label{dl2s-1}\ee

\no where we have the coefficients

\be C_j = \frac{(s-2j)(s-2j-1)}2 c_{j-1} + 2j(s-j)c_j \qquad ; \qquad j=1,2,\cdots,
\left\lbrack \frac{s-1}2\right\rbrack \quad . \label{Cjfix} \ee

\no Consequently, higher spin reparametrizations and Weyl invariance $C_j=0$ completely fixes $\lsm$ and the Cotton tensor up to an overall constant, i.e.,

\be c_j = \frac{(-1)^j (s-j-1)!}{4^j j! \, (s-2j-1)!} c_0 \qquad ; \qquad j=0,1,2,\cdots,
\left\lbrack \frac{s-1}2\right\rbrack \quad . \label{cj} \ee

\no Comparing (\ref{lsm}) with (\ref{l2s-1}) we have a closed formula for the Cotton tensor,

\be C_{\mu_1 \cdots \mu_s} = \Box^{s-1}E_{(\mu_1}^{\quad\rho} \sum_{j=0}^{\left\lbrack \frac{s-1}2\right\rbrack} \tilde{c}_j\, \left\lbrack \hth^{j}\theta^{s-1-j}\, h\right\rbrack_{\rho\mu_2 \cdots \mu_s )} \quad , \label{cotton} \ee

\no where $\tilde{c}_j=c_j$ at $c_0=1/s$. For\footnote{It is understood that all equalities involving Lagrangians in the present work hold under
space-time integrals.} the first four integer spins $\lsm$ becomes\footnote{Notice that in the spin-1 case we have replaced $h_{\mu}$ by the usual notation $A_{\mu}$ for the electromagnetic potential in order to avoid confusion with the spin-3 vector trace $h_{\mu}=\eta^{\rho\gamma}h_{\rho\gamma\mu}$.}

\bea \mathcal{L}^{(1)}_{1}=c_{0}\,A_{\mu}E^{\mu\nu}A_{\nu} \quad ; \quad \mathcal{L}^{(2)}_{3}=c_{0}\,h_{\mu_{1}\mu_{2}}\square{E}^{\mu_{1}\alpha_{1}}\theta^{\mu_{2}\alpha_{2}}h_{\alpha_{1}\alpha_{2}}\label{cs}\eea

\bea \mathcal{L}^{(3)}_{5}=c_{0}\,h_{\mu_{1}\mu_{2}\mu_{3}}\square^{2}{E}^{\mu_{1}\alpha_{1}}\Big[\theta^{\mu_{2}\alpha_{2}}\theta^{\mu_{3}\alpha_{3}}-\frac{1}{4}\theta^{\mu_{2}\mu_{3}}\theta^{\alpha_{2}\alpha_{3}}\Big]h_{\alpha_{1}\alpha_{2}\alpha_{3}}\label{cs3}\eea

\bea \mathcal{L}^{(4)}_{7}=c_{0}\,h_{\mu_{1}\mu_{2}\mu_{3}\mu_{4}}\square^{3}{E}^{\mu_{1}\alpha_{1}}\Big[\theta^{\mu_{2}\alpha_{2}}\theta^{\mu_{3}\alpha_{3}}\theta^{\mu_{4}\alpha_{4}}-\frac{1}{2}\theta^{\mu_{2}\mu_{3}}\theta^{\alpha_{2}\alpha_{3}}\theta^{\mu_{4}\alpha_{4}}\Big]h_{\alpha_{1}\alpha_{2}\alpha_{3}\alpha_{4}}\label{cs4}\eea

\no Notice that all non contracted indices $\mu_1 \cdots \mu_s$ on the right hand side of  (\ref{cotton}) come from transverse operators $E_{\mu\nu}$ and $\theta_{\mu\nu}$, so the transverse property of the Cotton tensor is explicit as in the case of the formulas given in
\cite{buch} in terms of spin projection operators\footnote{For an earlier connection between the Cotton tensor and projection operators in the $s=2$ case see \cite{unitarity}.}. Although we do not have a general proof we believe that (\ref{cotton}) does agree with previous formulae given in \cite{pt}, see also \cite{leonard}, for arbitrary integer spin.

In order to deduce $\lst$, instead of taking the symmetrized curl of the Cotton tensor (\ref{cotton}) we find more convenient to repeat the same procedure used for $\lsm$. We start from  a parity even Ansatz explicitly invariant under higher spin reparametrizations with arbitrary coefficients $b_j$, with $j=0,1,\cdots,\left\lbrack \frac s2 \right\rbrack $.

\be \lst = h_{\mu_1\cdots \mu_s}\Box^{s}  \left\lbrack b_0 \theta^{s} + b_1 \hth^2 \theta^{s-2} + b_2 \hth^4 \theta^{s-4}+ \cdots\cdots\cdots \right\rbrack^{\mu_1\cdots \mu_s\, \nu_1\cdots \nu_s} h_{\nu_1\cdots \nu_s} , \label{l2s}\ee

\no Notice that any even number of $E$ operators can be traded into
$\theta$ operators according to (\ref{id1}). Requiring generalized Weyl symmetry  we obtain an unique solution up to an overall constant,

%
%
%
%

\be b_j = \frac{(-1)^j s(s-j-1)!}{4^j j! \, (s-2j)!} b_0 \qquad ; \qquad j=0,1,\cdots,
\left\lbrack \frac{s}2\right\rbrack \quad . \label{dj} \ee

\no The first four cases of $\lst$ are given by

\be \mathcal{L}^{(1)}_{2}=b_{0}\, A^{\mu}\Box \theta_{\mu\nu}A^{\nu} = - \, \frac{b_{0}}2\, F_{\mu\nu}^2  \quad , \label{max} \ee

\bea \mathcal{L}^{(2)}_{4}=b_{0}\,h_{\mu_{1}\mu_{2}}\square^{2}\Big[\theta^{\mu_{1}\alpha_{1}}\theta^{\mu_{2}\alpha_{2}}-\frac{1}{2}\theta^{\mu_{1}\mu_{2}}\theta^{\alpha_{1}\alpha_{2}}\Big]h_{\alpha_{1}\alpha_{2}}=4\, b_{0}\Big(R^{2}_{\mu\nu}-\frac{3}{8}R^{2}\Big)_{hh} \quad , \label{kterm} \eea

\bea \mathcal{L}^{(3)}_{6}=b_{0}\,h_{\mu_{1}\mu_{2}\mu_{3}}\square^{3}\Big[\theta^{\mu_{1}\alpha_{1}}\theta^{\mu_{2}\alpha_{2}}\theta^{\mu_{3}\alpha_{3}}-\frac{3}{4}\theta^{\mu_{1}\mu_{2}}\theta^{\alpha_{1}\alpha_{2}}\theta^{\mu_{3}\alpha_{3}}\Big]h_{\alpha_{1}\alpha_{2}\alpha_{3}}\label{top3}\eea

\be \mathcal{L}^{(4)}_{8}=b_{0}\,h_{\mu_{1}\cdots\mu_{4}}\square^{4}\Big[\theta^{\mu_{1}\alpha_{1}}\theta^{\mu_{2}\alpha_{2}}\theta^{\mu_{3}\alpha_{3}}\theta^{\mu_{4}\alpha_{4}}-\theta^{\mu_{1}\mu_{2}}\theta^{\alpha_{1}\alpha_{2}}\theta^{\mu_{3}\alpha_{3}}\theta^{\mu_{4}\alpha_{4}} +\frac{\theta^{\mu_{1}\mu_{2}}\theta^{\mu_{3}\mu_{4}}\theta^{\alpha_{1}\alpha_{2}}\theta^{\alpha_{3}\alpha_{4}}}{8}\Big]h_{\alpha_{1}\cdots\alpha_{4}} \nn \\  \label{top4}\ee

\no  In the spin-1 case we have the Maxwell theory while in the spin-2 case we recognize the linearized K-term of the NMG theory \cite{bht}. In the next section we work out the spin-1 and spin-2 cases in terms of  appropriate gauge invariant field combinations as a preparation for the respective spin-3 and spin-4 cases since they turn out to have the same respective canonical structure.

\section{Spin-1 in terms of gauge invariants}

Before we start we stress that in the present section and throughout this work, the notation $i_{2s}$ and $i_{2s-1}$  stands for local invariants under the gauge transformation (\ref{gt1}) which are obtained via derivatives of order $2s$ and $2s-1$, respectively, of a rank-s fundamental field. There is in general a subset of the gauge transformations (\ref{gt1})
which leaves the lowest order Lagrangian $\lsb$ invariant.  Correspondingly we have the barred local invariants $\oi_{2s-1}$ and $\oi_{2s-2}$ of order $2s-1$ and $2s-2$ respectively. Notice that the same symbol may
represents different quantities in different spin-s section, see for instance (\ref{i1i2s1}) and (\ref{i4s2}). However, they both stand for an invariant built out of $2s$ derivatives of a rank-s tensor. Since each section deals only with one value of $s$, there will be hopefully no confusion.

Now let us start with the simplest spin-1 case where the gauge transformations (\ref{gt1}) become the usual $U(1)$ symmetry,

\be \delta \, A_{\mu} = \p_{\mu} \Lambda \quad . \label{gt2} \ee

\no One of the three equations (\ref{gt2}) can be used to eliminate the gauge parameter $\Lambda$ in terms of variations of the gauge field\footnote{Throughout this work, $i,j,k=1,2$ and henceforth we use  $\p_0 f$ and $\dot{f}$ equivalently.}, $\Lambda = \p_j\,\delta A_j/\nabla^2$, plugging back in (\ref{gt2}) we derive two local gauge invariants
$\delta \, i_{2s}=0=\delta \, i_{2s-1}$ connected with the electric and magnetic fields:

\be i_{2s} = \nabla^2 A_0 - \p_0(\p_k A_k) = \vec{\nabla} \cdot \vec{E}  \quad ; \quad i_{2s-1} = \hp_j A_j  = B \quad . \label{i1i2s1} \ee

\no We follow the notation of \cite{bht_hd} where

\be \hat{\p}_j = \epsilon_{jk}\p_k   \quad ; \quad \hat{\p}_i
\hat{\p}_j + \p_i\p_j = \nabla^2 \delta_{ij}  \quad ; \quad
 \hat{\p}_i\p_j - \hat{\p}_j\p_i = \nabla^2 \epsilon_{ij} \quad ; \quad \hat{\p}_i\hat{\p}_i = \p_j\p_j = \nabla^2 \label{hat}\ee

\no Introducing the so called helicity decomposition and redefining the gauge invariants  we have

\be A_0 = \rho \quad ; \quad A_j = \p_j\Gamma+\hp_j\gamma \quad , \label{amu}\ee

\be (\tp,\md) = (\tpi,\mdi)/\nabla^2 = (\rho - \dot{\Gamma},\gamma ) \quad . \label{I1I2s1} \ee

\no In all cases $s=1,2,3,4$ we will be able to write down the Lagrangians $\lst$ and $\lsm$ in terms of only two gauge invariants $\left(I_{2s}^{(s)},I_{2s-1}^{(s)}\right)$ and $\lsb$ in terms of those two and an extra one. In the spin-1 case the Maxwell and the Chern-Simons terms become

\bea \lst &=& b_{0}\, A^{\mu}\Box \theta_{\mu\nu}A^{\nu} =
b_0 \, \left\lbrack \tp(-\ns )\tp + \md(-\ns \Box )\md \right\rbrack \quad , \label{max} \\
\lsm &=& c_{0}\,A_{\mu}E^{\mu\nu}A_{\nu} = - 2\, c_{0}\, \md \, \nabla^2 \tp \quad .
\label{cs1} \eea

\no Since the two invariants can be treated as independent fields $(\tp,\md)=(\rho - \dot{\Gamma},\gamma ) \equiv (\orho ,\gamma )$, it is clear that the Abelian Chern-Simons term (\ref{cs1}) has no particle content\footnote{Throughout this work we assume vanishing fields at infinity. The Laplacian $\ns $ has only negative eigenvalues such that the frequently appearing operators  $\ns$ and $m^2-\ns$ have an empty kernel.} (topological term). We can combine (\ref{max}) with (\ref{cs1}) in order to produce the topologically massive Chern-Simons theory \cite{djt}. For future comparison with the spin-3 case we write it down with the choice $c_0= -m\, b_0$,

\bea \lsdt &=& b_0 \Big[ A^{\mu}\Box \theta_{\mu\nu}A^{\nu} - m\, A_{\mu}E^{\mu\nu}A_{\nu}\Big] \label{lmcs1} \\ &=&  b_0 \left\lbrace I_m(-\ns) I_m +   \md\left\lbrack -\ns (\Box - m^2 )\right\rbrack \md \right\rbrace\label{mcs}
\eea

\no where $I_m=\tp-m\, \md= \rho - \dot{\Gamma}-m\, \gamma \equiv \tilde{\rho}$  is the non propagating gauge invariant  while the transverse mode $\md = \gamma$ is the propagating one.

In the spin-1 case the Cotton tensor becomes a vector, the dual of the field strength: $C_{\mu} = E_{\mu\nu}A^{\nu}$. So, according to (\ref{dtensor}) we have $D_{\mu}=A_{\mu}$. Thus, $\lsb$ becomes the usual Proca mass term $A_{\mu}A^{\mu}$ which has of course no particle content and no gauge symmetry. So, there are no barred (residual) gauge transformations at all. However, in order to use a unified notation regarding the spin-3 case where non trivial barred gauge symmetries do in fact exist, we keep calling each of the components of the vector field $A_{\mu}$ a barred gauge invariant and keep using barred notation for some of the invariants.  The reader can check that the following expressions, which will have a spin-3 counterpart, hold true

\bea \lsb &=& d_0 \, A_{\mu}A^{\mu} = -d_0\left( \md \, \ns \, \md +  \mdb \ns \mdb + \bt^2  \right) \label{procam1} \\
&=& -d_0\left( \md \, \ns \, \md + \tp^2 + 2 \tp\, \dot{\oI}_{2s-1} + \mdb \Box \,\mdb \right) \label{procam2} \eea

\no where, recall $(\tp,\md)=(\rho - \dot{\Gamma},\gamma )$,

\be \mdb \equiv \p_j A_j/\ns = \Gamma  \quad ; \quad  \bt \equiv \tp + \dot{\overline{I}}_{2s-1} = A_0 = \rho \quad . \label{i1s1bar} \ee

\no We can choose  $(c_0,d_0)=(m\, b_0,-m^2b_0)$ and combine (\ref{cs1}) with (\ref{procam2}) in order to produce the first-order self-dual model $\lsdm$ of \cite{tpn}

\bea \lsdm &=& m\,b_0 \left( A_{\mu}E^{\mu\nu}A_{\nu} - m\, A_{\mu}A^{\mu}\right) \label{tpn} \\
& = & b_0 \left\lbrack I_{m}\,\ns\, I_{m} + \tilde{I}_{2s}(m^2-\ns) \tilde{I}_{2s}  + m^2  \mdb \frac{(-\ns ) (\Box - m^2)}{m^2-\ns}\mdb\right\rbrack \label{sdm1}
\eea

\no where

\be \tilde{I}_{2s}  \equiv \tp +
\frac{m^2}{m^2-\ns} \,\dot{\oI}_{2s-1}= \rho + \frac{\ns}{m^2-\ns}\, \dot{\Gamma} \label{i2st} \ee

\no The three gauge invariants can be treated as independent fields $(I_{m},\tilde{I}_{2s},\mdb)\equiv (-m\,\tilde{\gamma},\tilde{\rho}, \Gamma)$. Although (\ref{tpn}) is known \cite{dj} to be  dual do (\ref{lmcs1}), it contains one extra non propagating gauge invariant and the propagating mode is now the longitudinal component ($\Gamma $) instead of the transverse one ($\gamma $).

We finish this section building up the Maxwell-Proca theory
(parity doublet).
Since the Chern-Simons terms in (\ref{lmcs1}) and (\ref{tpn}) have opposite signs, we can simply add them and obtain

\bea \ldt &=& -\frac 14 F_{\mu\nu}^2 -\frac{m^2}2 A_{\mu} \label{mp1} \\
&=&   \md \frac{\left\lbrack -\ns (\Box - m^2 )\right\rbrack}2 \md
+ \frac {m^2}2  \mdb \frac{(-\ns ) (\Box - m^2)}{m^2-\ns}\mdb + \tilde{I}_{2s}\frac{(m^2-\ns)}2 \tilde{I}_{2s} \quad .  \label{mp2} \eea

\no Now we have both $\Gamma$ and $\gamma$ as propagating modes and the gauge invariant $\tilde{I}_{2s}$, connected with the electrostatic potential $A_0$, is the non propagating one. In all higher spin cases $s=2,3,4$ we will be able to write the doublet model in terms of two propagating and one non propagating gauge invariant.

\section{Spin-2 in terms of gauge invariants}

In the spin-2 case the gauge transformations become the usual linearized reparametrizations (diff) plus Weyl that we call WDiff,

\be \delta\, h_{\mu\nu} = \p_{\mu}\Lambda_{\nu} + \p_{\nu}\Lambda_{\mu} + \eta_{\mu\nu}\, \psi \equiv \delta_{\Lambda}h_{\mu\nu} + \delta_{\psi}h_{\mu\nu} \label{gt3} \ee

\no In (\ref{gt3}) we have 6 equations and 4 independent gauge parameters, consequently they give rise to 6-4=2 gauge invariants $(\tp,\md)$. This is all we need to describe $\lst$ and $\lsm$. It is instructive to do it in two steps. First we derive the 6-3=3 diff invariants. Using the decomposition
$\Lambda_0 = A$ , $\Lambda_j = \hp_j B + \p_j C $  in $\delta_{\Lambda}h_{\mu\nu}$ we find

\be A = \frac{\p_j\delta_{\Lambda}h_{0j}}{\ns } - \frac{\p_i\p_j\delta_{\Lambda}\dot{h}_{ij}}{2\nabla^4} \quad ; \quad B = \frac{\p_i\hp_j\delta_{\Lambda}h_{ij}}{\nabla^4} \quad ; \quad C = \frac{\p_i\p_j\delta_{\Lambda}h_{ij}}{2\nabla^4} \label{ABC} \ee

\no Substituting back in $\delta_{\Lambda}h_{\mu\nu}$ we derive 3 local diff invariants $\delta_{\Lambda}\oi_{2s}=0=\delta_{\Lambda}\oi_{2s-1}=\delta_{\Lambda}\oi_{2s-2}$,

\bea \oi_{2s-2} &=& \hp_i\hp_j h_{ij} \quad ; \quad \oi_{2s-1}= \ns \, \hp_j\, h_{0j} - \p_k \, \hp_j\, \dot{h}_{kj} \label{i2bi3bs2} \\ \oi_{2s} &=& \nabla^4\, h_{00} - 2\, \nabla^2 \p_j\, \dot{h}_{0j} + \p_i\p_j \ddot{h}_{ij} \label{i4b} \eea

\no The fact that we have only 3 independent local diff invariants is in agreement with the   Riemannian geometry, since in $D=2+1$ the Riemann tensor $R_{\mu\nu\alpha\beta}$ is proportional to the Ricci tensor $R_{\mu\nu}$ which has in principle 6 components but due to the Bianchi identity $\nabla^{\mu}R_{\mu\nu} = \nabla_{\nu}R/2$ only three of them are independent. Indeed, one can check that all six components of the linearized tensor $R_{\mu\nu}^L$  can be written in terms of space time derivatives of $(\oi_{2s-2},\oi_{2s-1},\oi_{2s})$.

Secondly, back to the WDiff symmetry, since
$\delta_{\psi}(\oi_{2s-2},\oi_{2s-1},\oi_{2s})=(\ns \, \psi,0,-\ns\, \Box\psi )$, we have two WDiff invariants,

\bea i_{2s-1}&\equiv &\oi_{2s-1} = \ns \, \hp_j\, h_{0j} - \p_k \, \hp_j\, \dot{h}_{kj} \label{i3s2} \\
i_{2s} &= &\oi_{2s} + \Box \, \oi_{2s-2} = \nabla^4\, h_{00} - 2\, \nabla^2 \p_j\, \dot{h}_{0j} + \p_i\p_j \ddot{h}_{ij} + \Box \, \hp_i\hp_j h_{ij} \quad .  \label{i4s2}\eea

 Using the helicity decomposition

\be h_{00} = \rho \quad ; \quad h_{0j} = \p_j\Gamma + \hp_j\gamma \quad ; \quad h_{ij}= \p_i\p_j \chi + (\p_i\hp_j + \hp_i\p_j)\theta + \hp_i\hp_j \varphi \label{hd2} \ee

\no and redefining the gauge invariants we have

\be (\oI_{2s-2},I_{2s-1}, I_{2s}) = (\frac 1{\nabla^4})(\oi_{2s-2},\oi_{2s-1},\oi_{2s}) = (\varphi,\gamma-\dot{\theta},\rho-2\dot{\Gamma}+ \ddot{\chi} + \Box\varphi ) \label{i2bi3i4} \ee

\no The fourth order linearized K-term of the NMG theory \cite{bht} and the third order linearized gravitational Chern-Simons term, see \cite{djt}, of the TMG can be written in terms of the two WDiff invariants

\bea \lst &=&  b_{0}\,h_{\mu_{1}\mu_{2}}\square^{2}(\theta^{\mu_{1}\alpha_{1}}\theta^{\mu_{2}\alpha_{2}}-\frac{\theta^{\mu_{1}\mu_{2}}\theta^{\alpha_{1}\alpha_{2}}}2) h_{\alpha_{1}\alpha_{2}} =
\frac{b_0}2 \, \left\lbrack \tp\,\nabla^4\,\tp + 4\md\,\nabla^4\, \Box \, \md \right\rbrack \quad , \label{lst2} \\
\lsm &=& c_{0}\,h_{\mu_{1}\mu_{2}}\square{E}^{\mu_{1}\alpha_{1}}\theta^{\mu_{2}\alpha_{2}}h_{\alpha_{1}\alpha_{2}} = -2\, c_{0}\, \md \, \nabla^4 \tp \quad .
\label{lsm2} \eea

From (\ref{i2bi3i4}) we see that the WDiff invariants may be considered as  independent fields $(I_{2s-1}, I_{2s})\equiv (\tilde{\gamma},\tilde{\rho})$,
thus we have a massless mode in (\ref{lst2}) and no particle content in (\ref{lsm2}). They can be combined together following (\ref{lsdt}), in order to produce the linearized version of the ``New Topologically Massive Gravity'' (NTMG) of \cite{sd4,andringa}, choosing  $c_0=-m\, b_0$ we have

\be \lsdt = \lst + \lsm  =  2\, b_0 \, \left\lbrace \frac 14 I_{2m}\, \nabla^4\, I_{2m} +  \md\left\lbrack \nabla^4 (\Box - m^2 )\right\rbrack \md\right\rbrace \label{ntmg}
\ee

\no where $I_{2m}=\tp+2m\, \md= \rho - \dot{\Gamma}+ \ddot{\chi} + \Box\varphi +2\, m (\gamma -\dot{\theta}) )$ and  $\md = \gamma -\dot{\theta}$.
Note that the change of variables $(I_{2m},\md)=(\tilde{\rho},\tilde{\gamma})$  has a trivial Jacobian. So we are able to check that the fourth order model (\ref{ntmg}) is free of ghosts and contain only one propagating  massive mode $\tilde{\gamma}$   plus the non propagating field $\tilde{\rho}$.

Now in order to construct $\lsb$ we need to find a D-tensor satisfying (\ref{dtensor})  and  such that $\lsb = d_0 \, h_{\mu\nu}D^{\mu\nu}$ has no particle content. In the spin-2 case the Cotton tensor is a symmetric rank-2 tensor, see (\ref{cs}). So with  $c_0=2$ so we need to solve the equation:

\be C_{\mu\nu} =  \left\lbrack E_{(\mu}^{\!\quad\alpha} \Box \theta_{\!\quad\nu )}^{\beta} h_{\alpha\beta}\right\rbrack =  E_{(\mu}^{\!\quad\rho}D_{\rho\nu)} \quad . \label{cmn} \ee

\no The D-tensor must be symmetric and of second order in derivatives. After a rather general Ansatz $D_{\mu\nu} \sim (\p^2 h)_{(\mu\nu )}$ we arrive, up to trivial redefinitions
$h_{\mu\nu}\to h_{\mu\nu} + \lambda\, \eta_{\mu\nu} h \,\, , (\lambda \ne - 1/3)$, at a general solution in terms of two arbitrary real parameters $(a,b)$, without loss of generality,

\be D_{\mu\nu} = \Box\, h_{\mu\nu} - \p_{(\mu}\p^{\rho}h_{\rho\nu )} + a\, \p_{\mu}\p_{\nu}\, h + (a \,\p^{\alpha}\p^{\beta}h_{\alpha\beta} - b\, \Box \, h)\eta_{\mu\nu} \label{dmn} \ee

\no Under the WDiff gauge transformations (\ref{gt1}) we have

\be \delta D_{\mu\nu} = \eta_{\mu\nu}\, \Box \, \left\lbrack 2(a-b)\p \cdot \Lambda + (a-3b+1)\, \psi \right\rbrack + \p_{\mu}\p_{\nu} \left\lbrack 2(a-1)\p\cdot\Lambda + 3\, a \psi\right\rbrack \label{ddmn} \ee

\no The WDiff invariance of the Cotton tensor $\delta C_{\mu\nu}= E_{(\mu}^{\!\quad\rho}\delta D_{\rho\nu)}=0$ for arbitrary values of $(a,b)$ follows simply from the tensor structure of $\delta D_{\mu\nu}$. Notice that $\delta D_{\mu\nu}=0$ for transverse diffeomorphisms (TDiff) : $(\p\cdot\Lambda,\psi)=(0,0)$. Thus, $\lsb = d_0 \, h_{\mu\nu}D^{\mu\nu}$ becomes the TDiff theory in $D=2+1$, see \cite{blas},

 \be \lsb (a,b) =   d_0 \left\lbrack -\del_\mu h^{\alpha\beta}\del^\mu h_{\alpha\beta}+ 2\,\del^\mu h^{\alpha\beta} \del_{\alpha}h_{\mu\beta}-2\,a\;\del^\mu h \del^\nu h_{\mu\nu}+b\,\del_\mu h \del^\mu h\right\rbrack
\label{lab} \ee

 Now we point out an interesting connection with massles spin-2 particles in $D=3+1$. Namely, it is known \cite{van} that TDiff is the minimal symmetry required for describing helicity $\pm 2$ particles in $D=3+1$ in terms of a symmetric rank-2 tensor $h_{\mu\nu}$. The general solution (\ref{ddmn}) seems to confirm that this is true also in $D=2+1$, since we can combine $\lsb (a,b)$ with the third order Chern-Simons term (\ref{lsm2}) and build up a third order  model that generalizes the linearized topologically massive gravity and contains helicity $2\vert m \vert /m$ particles. Such model can be non linearly extended to a topologically massive TDiff gravity since the metric determinant behaves as a scalar field under TDiff.
 Although there are descriptions of helicity $\pm 2$ in $D=2+1$ even  without gauge symmetry, see \cite{ak},  those models require auxiliary fields besides the symmetric rank-2 tensor $h_{\mu\nu}$. The FP conditions are enforced via second class constraints instead of local symmetries.

It is important to stress that (\ref{lab}) describes in general a massless scalar field in $D=2+1$. We can only have an empty spectrum if we enlarge the TDiff symmetry either to unconstrained linearized diffeomorphisms (Diff) by fixing\footnote{Up to trivial shifts $h_{\mu\nu} \to h_{\mu\nu} + \lambda \eta_{\mu\nu} h $ with $\lambda \ne -1/3$.} $(a,b)=(1,1)$ or to WTDiff (Weyl plus TDiff) by choosing $(a,b)=(2/3,5/9)$. The second case has been investigated in \cite{ghosh} and corresponds to the linearized version of unimodular gravity, its higher spin analogue, of second order in derivatives, has been investigated in \cite{sv}. A possible generalization of order $2s-2$ in $D=2+1$ will be studied elsewhere \cite{unp} from the point of view of gauge invariants. Here we only work with the linearized Einstein-Hilbert theory $\lsb = \lsb (1,1)$. In terms of the Diff invariants (\ref{i2bi3i4}) we have, see also \cite{djt},

\be \lsb =  {\cal L}_{LEH} = 2\, d_0 \left\lbrack \md \, \nabla^4 \, \md + \tp \, \nabla^4 \, \bt - \bt \, \nabla^4 \, \Box \bt \right\rbrack \quad . \label{leh} \ee

\no Since we can redefine $(\oI_{2s-2},I_{2s-1}, I_{2s}) =  (\varphi,\tilde{\gamma},\tilde{\rho})$, the equations of motion for those fields lead to the triviality of the EH theory in $D=2+1$: $\varphi=0=\tilde{\gamma}=\tilde{\rho}$.

Following (\ref{lsdm}) we can combine the Einstein-Hilbert theory (\ref{leh})  with the third order Chern-Simons term (\ref{lsm2})   and build up the linearized version of TMG, choosing  $(c_0,d_0) = (-m\, b_0,-m^2 b_0) $,

\bea \lsdm &=& \lsm + \lsb ={\cal L}_{TMG} \nn\\ &=& 2\, m^2\, b_0 \left\lbrack \bt \, \nabla^4 \, \Box \, \bt - \md \, \nabla^4 \md - \tp \, \nabla^4 \left(\bt + \frac{\md}m \right)\right\rbrack \quad . \label{ltmg} \eea

\no Since the Lagrangian is linear on  $\tp$ we have the functional constraint $\md = -m\, \bt$ which leads to $\lsdm = 2\, m^2\, b_0 \left\lbrack \bt \, \nabla^4 \, (\Box - m^2) \bt \right\rbrack $ confirming that we have one physical massive mode content without ghosts. Finally we simply add (\ref{ntmg}) and (\ref{ltmg}) in order to produce the NMG parity doublet,

\be \ldt = {\cal L}_{LNMG} = 2 b_0 \left\lbrack \md \, \nabla^4 (\Box -m^2)  \md + m^2 \bt \, \nabla^4 (\Box -m^2) \, \bt + \oI_{2s}\frac{\nabla^4}4\oI_{2s} \right\rbrack  \label{lnmg} \ee

\no where $ \oI_{2s} =\tp -2\, m^2 \bt $. Since $(\oI_{2s},\md,\bt)$ are independent degrees of freedom we confirm the doublet ghost free content of the linearized NMG in a gauge invariant way.

\section{Spin-3}

In the rank-3 case the WDiff transformations (\ref{gt1}) become the following 10 equations

\be \delta h_{\mu\nu\alpha} = \p_{(\mu}\Lambda_{ \nu\alpha)} + \eta_{(\mu\nu} \psi_{\alpha )} \quad ,\label{gt4}\ee

\no At first sight we have $9$ gauge parameters on the right hand side
of (\ref{gt4}), however there are only 8 independent ones due to the redundancy $\delta (\Lambda_{\nu\alpha},\psi_{\alpha}) = (\eta_{\nu\alpha}\phi,-\p_{\alpha}\phi)$. An equivalent counting, valid for arbitrary spin $s\ge 3$ as we will see in section 8,  is to consider, without loss of generality, that we can replace arbitrary Diff in (\ref{gt4}) by traceless  Diff $\oL_{\nu\alpha}$ ($\eta^{\nu\alpha}\oL_{\nu\alpha}=0$). No redundancy is left in this case.

From (\ref{gt4}) we have 10-8=2 gauge invariants. In practice we can decompose $\psi_{\mu}$ and $\oL_{\mu\nu}$
according to formulae similar to (\ref{amu}) and (\ref{hd2}) and find out explicit expressions for the 8 independent gauge parameters in terms  of $\delta h_{\mu\nu\alpha}$, recall that $\oL_{00}=\oL_{jj}$ . Plugging back in (\ref{gt4}), after some work, we have $\delta i_{2s} = 0=\delta i_{2s-1}$ where the 6th and 5th order local WDiff invariants are

\be i_{2s} = \nabla^6 h_{000} - 3\nabla^4 \p_j\p_0\, h_{00j} + 3\nabla^2
\p_j\p_k\p_{0}^2 \, h_{0jk} - \p_i\p_j\p_k \p_{0}^3 h_{ijk}+3\, \Box \,(\nabla^2 \hp_j\hp_k \,h_{0jk} - \hp_j\hp_k\p_l \p_0 \,h_{jkl} ) \label{i6}\ee

\be i_{2s-1} = 3 \left( \p_i\p_j\hp_k\p_0^2 h_{ijk} - 2 \nabla^2 \p_0\p_j\hp_k \,h_{0jk} + \nabla^4 \hp_j \,h_{00j} \right) + \Box \,\hp_j\hp_k\hp_l  \,h_{jkl}  \label{i5}\ee

\no Introducing the helicity decomposition

\bea h_{000} &=& \rho \quad ; \quad h_{00j} = \p_j\,\Gamma + \hp_j\, \gamma  \label{hmna1}\\ h_{0jk} &=& \hp_j\hp_k \phi_1 + \p_j\p_k \phi_2 + \hp_{(j}\p_{k)}\, \phi_3 \label{hmna2} \\
h_{jkl} &=& \hp_j\hp_k\hp_l \, \psi_1 + \hp_{(j}\hp_k\p_{l)} \,\psi_2
+ \p_{(j}\p_k\hp_{l)} \,\psi_3  + \p_j\p_k\p_l \,\psi_4 \quad , \label{hmna3} \eea

\no and redefining the invariants we have

\bea \tp \equiv i_{2s}/\nabla^6 &=& \rho - 3\, \p_0\, \Gamma + 3\, \Box\, \phi_1 + 3\, \p_0^2\phi_2 - \p_0^3\psi_4 - 3\, \Box\, \p_0\, \psi_2  \label{I6} \\ \md \equiv i_{2s-1}/\nabla^6 &=& 3\, \gamma - 6\, \p_0\, \phi_3 + 3\, \p_0^2 \psi_3 + \Box\, \psi_1 \quad  . \label{I5} \eea

\no The next step is to write down $\lst$ and $\lsm$ given in (\ref{top3}) and (\ref{cs3}) respectively in terms of the gauge invariants (\ref{I6}) and (\ref{I5}). This is much more complicate than in the previous $s=1,2$ cases
where the explicit substitution of the helicity decomposition could be easily carried out. Now we use a short cut. Namely, we suppose that in both cases the searched Lagrangian has the form

\be {\cal L} = \tp \, \hat{A} \, \tp + \md \, \hat{B} \, \md + \tp \, \hat{C} \, \md \quad . \label{ABC} \ee

\no where $(\hat{A},\hat{B},\hat{C})$ are space time differential operators to be found. We restrict the decomposition (\ref{hmna1})-(\ref{hmna3}) to the smallest number of fields which allows us to find out the unknown differential operators\footnote{Alternatively, we believe that is possible to determine the operators $(\hat{A},\hat{B},\hat{C})$, up to an overall constant, by Lorentz invariance, mass dimension and locality as we have done in some examples.}. We have found that the most convenient choice is to  keep only $\psi_1$ and $\psi_2$. We are left only with spatial components of the fundamental field,

\be h_{ijk} = \hp_i\hp_j\hp_k \, \psi_1 + \hp_{(i}\hp_j\p_{k)} \,\psi_2  \quad . \label{psi13} \ee

\no Thus, we have

\bea \lst \!\!\!\! &=& b_0\,  h_{ijk}\,  \left\lbrack \theta^{im}\theta^{jn}\theta^{kp} - \frac 34 \theta^{ij}\theta^{mn}\theta^{kp} \right\rbrack \Box^3 h_{mnp} \label{l2s3a} \\
\!\!\!\! &=& b_0 \, \hp_i\hp_j\hp_k \, \psi_1 \left\lbrack \delta^{im}\delta^{jn}\delta^{kp} -\frac 34 \delta^{ij}\delta^{mn} \delta^{kp}\right\rbrack \Box^3\, \hp_m\hp_n\hp_p \psi_1
\nn \\ \!\!\!\!\!\! &+& \!\!\! b_0 \,  \hp_{(i}\hp_j\p_{k)} \,\psi_2 \left\lbrack \delta^{jn}\delta^{kp}(\Box\, \delta^{im}-3\, \p^i\p^m) - \frac 34 \delta^{ij}\delta^{mn}(\Box\, \delta^{kp} - \p^k\p^p)\right\rbrack \Box^2 \, \hp_{(m}\hp_n\p_{p)} \,\psi_2    \label{73b} \\
\!\!\!\! &=& b_0 \left\lbrack \Box\, \psi_1\left(\frac{-\nabla^6\Box}4 \right)\Box\, \psi_1 + (-3\, \Box\, \dot{\psi}_3)\left(\frac{-\nabla^6}4\right) (-3\, \Box\, \dot{\psi}_3)\right\rbrack \label{73c} \\ &=& b_0 \left\lbrack \md\left(\frac{-\nabla^6\Box}4 \right) \md + \tp \left(\frac{-\nabla^6}4\right) \tp  \right\rbrack \label{itop3} \eea

\no where $\Box \theta^{im} = \Box \delta^{im} - \p^i\p^m $ and from (\ref{I6}), (\ref{I5}) we have  $ (\tp,\md) = (- 3\, \Box\,  \dot{\psi}_2,\Box\, \psi_1)$. Due to the fact that there is always an odd (even) number of dual derivatives $\hp$ acting on $\psi_1$ ($\psi_2$) there are no cross terms $\psi_1 \times \psi_2$, they vanish due to  $\hp \cdot \p =0$. For the same reason we have dropped several derivatives in (\ref{73b}).  Notice that the two terms inside parenthesis in  (\ref{73b}) have exactly led to the double time derivatives required to produce the second term of (\ref{73c}) which is a non trivial check of (\ref{ABC}). Similarly, for the fifth order spin-3 Chern-Simons term, using (\ref{psi13}) again, we have

\be \lsm = c_0 \, h_{ijk}\, E^{im} \left\lbrack \theta^{jn}\theta^{kp}- \frac 14 \theta^{jk}\theta^{np} \right\rbrack \Box^2 h_{mnp} =  c_0 \, h_{ijk} \left\lbrack \theta^{jn}\theta^{kp}- \frac 14 \theta^{jk}\theta^{np} \right\rbrack \Box^2 \dot{h}^*_{i(np)} \, , \label{lsm3a} \ee

\no where we have used $E^{im} =\epsilon^{im}\p_0$ and defined

\be h^*_{i(np)} \equiv \epsilon^{im}h_{mnp} = - \p_i \hp_n\hp_p \, \psi_1 + \hp_i\hp_n\hp_p \psi_2 - \p_i\p_{(n}\hp_{p)}\, \psi_2 \quad . \label{hdual3}\ee

\no Notice that in $h_{ijk}$ we have an odd (even) number of dual derivatives $\hp$ acting on $\psi_1$ ($\psi_2$) while the opposite applies for $h^*_{i(np)}$, therefore only  cross terms $\psi_1 \times \psi_2$ show up in (\ref{lsm3a}) and we can neglect the last term of (\ref{hdual3}) due to $\p \cdot \hp =0$. Consequently,

\bea \lsm &=& c_0 \, [ \hp_i\hp_j\hp_k \, \psi_1 + \hp_{(i}\hp_j\p_{k)} \,\psi_2 ] \left\lbrack \delta^{jn}\delta^{kp}- \frac 14 \delta^{jk}\delta^{np} \right\rbrack  \Box^2 (- \p_i \hp_n\hp_p \, \psi_1 + \hp_i\hp_n\hp_p \psi_2) \nn \\
&=& 2\, c_0 \, \Box \psi_1 \, \nabla^6 (-3\,\Box \, \dot{\psi}_2) = 2\, c_0 \left( \md \, \frac{\nabla^6}4 \, \tp \right) \quad . \label{ics3} \eea

From (\ref{I6}) and (\ref{I5}) we see that we can  define, with a trivial Jacobian, the new fields $(\tp,\md)\equiv (\tilde{\rho},3\, \tilde{\gamma})$ such that  $\tp$ and $\md $ are two independent fields as in the previous spin-1 and spin-2 cases. So we can verify by comparing (\ref{itop3}) with (\ref{max}) and (\ref{lst2}) as well as (\ref{ics3}) with (\ref{cs1}) and (\ref{lsm2}), that the canonical structure of $\lst$, $\lsm$ remains the same up to irrelevant overall numerical factors and powers of $-\ns$ which can be absorbed in redefinitions of the constants $(b_0,c_0)$ and of the invariants $(\tp,\md)$ respectively. There is  no obstacle in building up the spin-3 sixth order self-dual model $\lsdt $, as originally suggested in \cite{bht_hd}. By combining $\lst$ and $\lsm$ with $c_0=-m\, b_0$ we have a self-dual model with the same form of (\ref{mcs}),

\bea \lsdt &=& b_{0}\,h_{\mu\nu\rho}\Big[\square(\theta^{\mu\alpha}\theta^{\nu\beta}\theta^{\rho\gamma}-\frac{3}{4}\theta^{\mu\nu}\theta^{\alpha\beta}\theta^{\mu\gamma}) -m\,{E}^{\mu\alpha}(\theta^{\nu\beta}\theta^{\rho\gamma}-\frac{1}{4}\theta^{\mu\nu}\theta^{\beta\gamma}) \Big]\square^2 h_{\alpha\beta\gamma} \label{lsd6} \\
&=& \frac{b_0}{4} \left\lbrace I_m(-\ns) I_m +   \md\left\lbrack -\ns (\Box - m^2 )\right\rbrack \md \right\rbrace \label{ilsd6}\eea

\no From (\ref{I6}) and (\ref{I5}) we see that we can redefine the fields such that $\md \equiv \orho $ and $I_m\equiv \tp+m\, \md \equiv 3\overline{\gamma} $. So the particle content of (\ref{lsd6}) corresponds to only one propagating massive mode.

We  move now to the investigation of the fourth order spin-3 Lagrangian $\lsb = b_0 h_{\mu\nu\alpha}\, D^{\mu\nu\alpha}$, preliminarly studied in \cite{sd5}. We need to find the symmetric D-tensor which solves the equation

\be C_{\mu\nu\alpha} = E_{(\mu}^{\!\quad \rho}D_{\rho\nu\alpha )} \quad . \label{cmna} \ee

\no where the spin-3 Cotton tensor can be obtained from (\ref{cs3}) or (\ref{cotton}). The D-tensor must be of fourth order in derivatives ($D_{\mu\nu\alpha} \sim (\hp^4h)_{(\mu\nu\alpha)}$). In the spin-2 case we have started from a general second order Ansatz $D_{\mu\nu} \sim (\hp^2h)_{(\mu\nu)}$ and required (\ref{cmn}). Alternatively, we could have obtained (\ref{ddmn}) by requiring instead that its variation under WDiff had the tensor structure $\delta \, D_{\mu\nu} =  \p_{\mu}\p_{\nu} F + \eta_{\mu\nu}\, \Box \, G  $. This guarantees the WDiff invariance of the spin-2 Cotton tensor. The Cotton tensor is uniquely determined by its local symmetry and order in derivatives. Since $F$ and $G$ must be linear functions of $\p \cdot \Lambda $ and $\psi$ the required tensor structure is equivalent to the TDiff  symmetry. The spin-3 and spin-4 cases are completely analogous. In the spin-3 case, the symmetry of the Cotton requires that under (\ref{gt4}) we have $\delta D_{\mu\nu\alpha} = \p_{\mu}\p_{\nu}\p_{\alpha} F + \Box\, \eta_{(\mu\nu}\p_{\alpha )}G $ where $F$ and $G$ are linear functions of $\Box\,\Lambda$, $\p^{\mu}\p^{\nu}\Lambda_{\mu\nu}$  and $\p^{\mu}\psi_{\mu}$. This is equivalent to demand

\be \odl \int d^3 x \, h_{\mu\nu\alpha} D^{\mu\nu\alpha} = 0 \quad . \label{dbar} \ee

\no where the $\odl$ gauge transformations correspond to WDiff with the 3 scalar restrictions:

\be \eta^{\mu\nu}\Lambda_{\mu\nu} \equiv \Lambda =0 = \p^{\mu}\p^{\nu}\Lambda_{\mu\nu} = \p^{\mu}\psi_{\mu} \quad . \label{r1} \ee

\no The general solution to (\ref{dbar}) is a two parameter family of Lagrangians,

\bea \lsb (f,g) &=& d_0\Big[ h_{\mu\nu\alpha}\, \Box^2 h^{\mu\nu\alpha}
- \frac 34 h_{\mu}\, \Box^2 h^{\mu} -3\, h_{\mu\nu\alpha}\, \Box \p^{\mu}\p_{\beta} h^{\beta\nu\alpha} + \frac 32 h_{\mu\nu\alpha}\, \Box \p^{\mu}\p^{\nu}h^{\alpha}   \nn \\
&+& \frac 94 h_{\mu\nu\alpha}\p^{\mu}\p^{\nu}\p_{\beta}\p_{\rho} h^{\beta\rho\alpha} + f \, h_{\mu}\,\Box \,\p^{\mu}\p^{\nu}h_{\nu} + g \,  h_{\mu\nu\alpha}\p^{\mu}\p^{\nu}\p^{\alpha}\p^{\beta}h_{\beta}\Big] \label{l4} \eea

\no where the parameters ($f,g$) are so far arbitrary.

The question is: which values of the parameters ($f,g$) render (\ref{l4}) an empty theory ? The more symmetry, the less content we have, so we must try to enlarge the $\odl$-symmetry  as much as possible. In the spin-2 case the TDiff symmetry associated with the restrictions $(\p^{\mu}\Lambda_{\mu},\psi)=(0,0)$ could be enlarged either to Diff ($\psi =0$) or to WTDiff ($\p^{\mu}\Lambda_{\mu}=0$). So the idea is to lift the restrictions (\ref{r1}) as much as possible. The reader can check that there is no solution for ($f,g$) if we try to keep only one of the three restrictions (\ref{r1}), but in case we keep two of them we have found some solutions. In \cite{unp} we will make a general analysis including the more complex spin-4 case. Here we stick to the case where the  $\odl$-symmetry
is enlarged to traceless spin-3 Diff plus transverse Weyl transformations (TWDiff),

\be \Lambda = 0 = \p^{\mu}\psi_{\mu} \quad \to \quad (f,g)=(\frac{21}{16},-\frac{9}4) \label{2p} \ee

\no correspondingly we define from (\ref{l4}) the fourth order spin-3 Lagrangian

\be {\cal L}_4 ^{(3)} \equiv {\cal L}_{2s-2}^{(s)} \left( \frac{21}{16},-\frac 94 \right) \quad {\rm at} \quad d_0 = -m^2 b_0 \quad . \label{l43} \ee

 The reason we choose (\ref{2p}) is twofold. First, it has already been analysed, in a fixed gauge,  in \cite{sd5} where it is shown to have an empty spectrum\footnote{The spin-3 fourth order Lagrangian (\ref{l43}) appeared in the literature, see \cite{mkr}, even before \cite{sd5}. We thank Prof. Karapet Mkrtchyan for bringing that reference to our knowledge.}. Secondly, there will be an analogous case for spin-4 as we will see in the next section.

%

Following our gauge invariant approach, due to the restrictions $(\Lambda,\p^{\mu}\psi_{\mu})=(0,0)$ we have 7 independent gauge parameters on the right hand side of the 10 equations (\ref{gt4}), thus we have 10-7=3 gauge invariants just like the spin-2 Einstein-Hilbert case and the spin-1 Proca mass term. By eliminating the 7 independent gauge parameters as functions of $\delta\, h_{\mu\nu\alpha} $ and plugging back in (\ref{gt4}) we obtain two fifth order invariants and a sixth order one,
$\delta \, i_{2s-1} = 0 = \delta \, \oi_{2s-1}= \delta\, i_{2s} $ where $(i_{2s},i_{2s-1})$ are the two invariants under unrestricted transformations (\ref{gt4}) given in (\ref{i6}) and (\ref{i5}) while

\bea \oi_{2s-1} &=& 3 \big[ \ns \hp_i\hp_j\p_k \, h_{ijk} + 2 \, \hp_i\hp_j\p_k \, \ddot{h}_{ijk}- 3 \, \ns\hp_j\hp_k \dot{h}_{0jk} \big] + 3\big[ \p_j\p_k \dot{h}_{0jk}  - \, \ns \p_j \, \dot{h}_{00j} \big] \nn \\
&+& \ns \p_i\p_j\p_k \, h_{ijk} -2\, \p_i\p_j\p_k \, \ddot{h}_{ijk} + \nabla^4 \dot{h}_{000} \quad . \label{i5b} \eea

\no After a convenient redefinition, in terms of helicity variables, we have

\be \mdb =\frac{\mdbi}{(-2\,\nabla^6)} = \Big[-3\, \Gamma + \frac{\dot{\rho}}{\ns} - 9 \dot{\phi}_1 + 3\, \dot{\phi}_2 + \ns \psi_4 - 2 \ddot{\psi}_4 + 3(\ns\psi_2 +2\, \ddot{\psi}_2 )\Big]/(-2) \label{I5b} \ee

As in (\ref{ABC}) we assume that $\lsb (21/16,-9/4) = \sum_{K,L} I_K \hat{O}_{KL} \, I_L$
where the sum run over the 3 invariants (\ref{I6}),(\ref{I5}) and (\ref{I5b}) while $\hat{O}_{KL}$ stands for a symmetric $3\times 3$ matrix differential operator to be found. We have followed a two steps procedure. In the first step we keep only $(\psi_1,\psi_4)\ne (0,0)$ in the helicity decomposition (\ref{hmna1})-(\ref{hmna3}) while in the second one we assume $(\psi_1,\psi_2)\ne (0,0)$  such that we respectively have

\bea h_{jkl}&=&\hp_j\hp_k\hp_l \,\psi_1 + \p_j\p_k\p_l \, \psi_4 \to (\tp,\md,\mdb) = (-\p_0^3\psi_4,\Box\, \psi_1, \ddot{\psi}_4-\nabla^2\psi_4/2 \,) \label{hjkl1}\\
 h_{ijk} &=& \hp_i\hp_j\hp_k \, \psi_1 + \hp_{(i}\hp_j\p_{k)} \,\psi_2  \to (\tp,\md,\mdb) = (- 3\, \Box\,  \dot{\psi}_2,\Box\, \psi_1,-3 (\ns \, \psi_2 + 2 \ddot{\psi}_2)/2) \nn\\\label{hjkl2} \eea

\no Direct substitution in ${\cal L}_4^{(3)}$ leads respectively to

\bea {\cal L}_I &=& -\psi_1 \nabla^6 \Box^2 \psi_1 + \psi_4 \nabla^6 \big[ - \Box^2 +(3/4)\nabla^2\Box \big] \psi_4 \quad , \label{l1} \\
{\cal L}_{II} &=& \Box\, \psi_1 (-\nabla^6)\Box\, \psi_1 + 9\, \psi_2 \,\Box(\ddot{\psi}_3 - \ns \psi_2/4) \label{l2} \eea

\no which uniquely determine the fourth order spin-3 Lagrangian, compare with (\ref{procam1}) and (\ref{procam2}),

\bea \lsb  &=&   -d_0\left( \md \, \nabla^6 \md +  \mdb \nabla^6 \mdb + \bt^2  \right) \label{lsb3I1} \\
&=& -d_0\left( \md \, \nabla^6 \, \md + \tp \, \nabla^4 \, \tp + 2 \tp\, \nabla^4\dot{\oI}_{2s-1} + \mdb \nabla^4\Box \,\mdb \right) \label{lsb3I2} \eea

\no where we have defined

\be \bt = \ns (\tp + \dot{\oI}_{2s-1}) = 2\, \ns \rho - \ddot{\rho}- 3\, \ns \dot{\Gamma} + 3\, \ns (\ddot{\phi}_1 + 2\, \ns \phi_1)+ 3\, \ns \ddot{\phi}_2 - 9 \, \nabla^4 \dot{\psi}_3 - \nabla^4 \dot{\psi}_2 \label{I4b} \ee

\no Notice from (\ref{i6}) and (\ref{i5b}) that all sixth order terms in the combination $\bt $ cancel out and we are left with at most four derivatives of the fundamental field
$h_{\mu\nu\alpha}$ which justifies the lower index. Now an important technical point must be stressed. In order to establish full analogy with the spin-1
case we should be able to treat $(\md,\mdb,\bt)$ as independent fields.
Although $\md \equiv 3\, \gamma $ decouples from $\mdb $ and $\bt$, due to the time derivatives on $\Gamma$ and $\rho$ in (\ref{I5b}) and (\ref{I4b})
 it is not obvious that both $\mdb$ and $\bt$ can be treated as basic independent fields. In order to prove it we first get rid of time derivatives in (\ref{I5b}) redefining $\Gamma$ via $\mdb \equiv 3\, \oGamma/2 $. After such redefinition (\ref{I4b}) still have terms of the type $\ddot{\rho}$ which can be eliminated  via $\ophi_1 \equiv \phi_1 - \frac{\rho}{6\ns}$. The final step is to redefine $\rho$ according to
 $\bt \equiv 3\,\ns \orho $. The reader can check that the triple change of variables $\overline{\Phi}_J=M_{JK}\Phi_K + G_J $ with $\Phi_J = (\Gamma,\phi_1,\rho)$ and $G_J$ independent of $\Phi_J$  is such that all derivatives cancel out in the Jacobian and we have $\det M = 1 $. Therefore, the fourth order theory given in (\ref{lsb3I1}) or (\ref{lsb3I2}), see also (\ref{l4}) with $(f,g)=(21/16,-9/4)$,  has no particle content, in agreement with the gauge fixed analysis of \cite{sd5}.

 Since the spin-3 Lagrangian (\ref{lsb3I2}) has exactly the same form of the Proca mass term (\ref{procam2}), similarly for the spin-1 (\ref{cs1}) and spin-3 (\ref{ics3}) Chern-Simon terms, we can follow the same steps leading to (\ref{sdm1}), with the choice $(c_0,d_0)=(m\, b_0, - m^2b_0)$,  and obtain the fifth-order spin-3 self-dual model suggested in \cite{sd5} in terms of gauge invariants,

 \bea \lsdm &=& m\,b_0 \,  h_{\mu\nu\rho}\,{E}^{\mu\alpha}(\theta^{\nu\beta}\theta^{\rho\gamma}-\frac{1}{4}\theta^{\mu\nu}\theta^{\beta\gamma}) \square^2 h_{\alpha\beta\gamma} + {\cal L}_4^{(3)} \label{sd5} \\
& = & \frac{b_0}{4} \left\lbrack I_{m}\,\ns\, I_{m} + \tilde{I}_{2s}(m^2-\ns) \tilde{I}_{2s}  + m^2  \mdb \frac{(-\ns ) (\Box - m^2)}{m^2-\ns}\mdb\right\rbrack \label{isd5}
\eea

\no where $I_m=\tp+m\, \md$,  $\tilde{I}_{2s}  \equiv \tp +
m^2 \dot{\oI}_{2s-1}/(m^2-\ns)$ and ${\cal L}_4^{(3)}$ is given in (\ref{l43}). Such model is the spin-3 analogue of the linearized TMG. Notice however that  it is not trivial to show that $(\tilde{I}_{2s},\mdb,I_m)$ are three independent degrees of freedom. First we notice that $\gamma$ only appears in $I_m$, thus the redefinition $I_m = 3\,m\, \ogamma$ does not affect $(\tilde{I}_{2s},\mdb)$. Next we redefine $\Gamma $ such that $\mdb \equiv -3\, \oGamma$, then we make $\ophi_1 \equiv \phi_1 - \frac{\rho}{6\ns}$ in order to get rid of time derivatives of $\rho$ in $\tilde{I}_{2s}$, finally we  redefine $\rho $ such that $\tilde{I}_{2s}=3\ns (m^2-\ns )\orho $. It turns out that the whole Jacobian is trivial.

The doublet model $\ldt$, i.e., the spin-3 analogue of NMG has been suggested in \cite{dsmb} where it was obtained via soldering  of two self-dual models of  opposite helicities $+3$ and $-3$ as given in (\ref{lsd6}) or in (\ref{sd5}). The same result can be obtained adding (\ref{lsd6}) and (\ref{sd5}) with $c_0=m\, b_0$, i.e.,

\bea \ldt &=&  b_{0}\, h_{\mu\nu\rho}\,\square^3(\theta^{\mu\alpha}\theta^{\nu\beta}\theta^{\rho\gamma}-\frac{3}{4}\theta^{\mu\nu}\theta^{\alpha\beta}\theta^{\mu\gamma})h_{\alpha\beta\gamma} + {\cal L}_4^{(3)}  \label{d6} \\
&=& \frac{b_0}4 \left\lbrace    \md\left\lbrack -\nabla^6 (\Box - m^2 )\right\rbrack \md + m^2  \mdb \frac{(-\nabla^6 ) (\Box - m^2)}{m^2-\ns}\mdb + \tilde{I}_{2s}\nabla^4(m^2-\ns) \tilde{I}_{2s} \right\rbrace \nn\\ \label{id6} \eea

\section{The spin-4 case}

In the rank-4 case the WDiff transformations (\ref{gt1}) equations
 correspond to 15 equations:

\be \delta h_{\mu\nu\alpha\beta} = \p_{(\mu}\Lambda_{ \nu\alpha\beta)} + \eta_{(\mu\nu} \psi_{\alpha\beta )} \quad .\label{gt5}\ee

\no By either considering  $\Lambda_{ \nu\alpha\beta}$ a traceless tensor $\eta^{\nu\alpha}\Lambda_{\nu\alpha\beta} \equiv \Lambda_{\beta}=0$ or taking into account the  vector redundancy $\delta (\Lambda_{\nu\alpha\beta},\psi_{\alpha\beta})=(\eta_{(\nu\alpha}\epsilon_{\beta )},-\p_{(\alpha}\epsilon_{\beta )})$  we see that (\ref{gt5}) leads to 15-13=2 WDiff local invariants of 8th and 7th order in derivatives, $\delta \, i_{2s} = 0 = \delta\, i_{2s-1}$,

\bea i_{2s} &=& \nabla^8 h_{0000} - 4\nabla^6 \p_j\p_0\, h_{00j} + 6\nabla^4
\p_j\p_k\p_{0}^2 \, h_{00jk} - 4\ns\p_i\p_j\p_k \p_{0}^3 h_{0ijk} + \p_i\p_j\p_k\p_l \p_{0}^4 h_{ijkl} \nn\\ & + & 6\,\Box\Big[  (\nabla^4 \hp_j\hp_k \,h_{00jk} - 2\, \hp_j\hp_k\p_l \p_0 \,h_{0jkl} + \hp_j\hp_k\p_l\p_m \p_0^2 \, h_{jklm} \Big] + \Box^2 \hp_j\hp_k\hp_l\hp_m  \, h_{jklm} \label{i8}\eea

\bea i_{2s-1} &=& \nabla^6 \hp_j \,h_{000j}  - 3 \nabla^4 \p_j\hp_k\p_0 \,h_{00jk} +3\,\ns \p_i\p_j\hp_k\p_0^2 h_{0ijk}-\p_i\p_j\p_k\hp_l\p_0^3\, h_{ijkl} \nn \\
& &\Box  \Big[ \ns \hp_i\hp_j\hp_k\, h_{0ijk} - \p_i\hp_j\hp_k\hp_l\p_0 h_{ijkl} \Big] \label{i7}\eea

\no After the helicity decomposition

\bea h_{0000} &=& \rho \quad ; \quad h_{000j} = \hp_j\, \gamma + \p_j\,\Gamma   \label{hmnab1}\\ h_{00jk} &=& \hp_j\hp_k \phi_1 + \hp_{(j}\p_{k)}\, \phi_2 + \p_j\p_k \phi_3  \label{hmnab2} \\
h_{0jkl} &=& \hp_j\hp_k\hp_l \, \psi_1 +  \hp_{(j}\hp_k\p_{l)} \,\psi_2
+ \p_{(j}\p_k\hp_{l)} \,\psi_3 + \p_j\p_k\p_l \,\psi_4 + \quad , \label{hmnab3}\\
h_{ijkl} &=& \hp_i\hp_j\hp_k\hp_l \, \beta_1 +  \hp_{(i}\hp_{j}\hp_k\p_{l)} \,\beta_2
+ \p_{(i}\p_{j}\hp_k\hp_{l)} \,\beta_3 + \hp_{(i}\p_j\p_k\p_{l)} \,\beta_4 +
\p_{i}\p_j\p_k\p_{l} \,\beta_5 \, , \label{hmnab4} \eea

\no and redefining the invariants we obtain

\bea \tp & &\equiv i_{2s}/\nabla^8 = \rho - 4\, \dot{\Gamma} + 6 \ddot{\phi}_3 + 6 \, \Box\, \phi_1 - 4\p_0^3\psi_4 - 12\, \Box\, \p_0\, \psi_2 + \p_0^4 \beta_5 + \Box^2 \beta_1 \label{I8} \\
\md & &\equiv i_{2s-1}/\nabla^8 =  \gamma + 3\, \ddot{\psi}_3 + \Box\, \psi_1 - \Box \, \dot{\beta}_2 - \p_0^3 \beta_4 \quad  . \label{I7} \eea

In order to write down the 8th order WDiff invariant Lagrangian (\ref{top4})
in terms of gauge invariants we assume that the only non vanishing fields  are $\beta_1$ and $\beta_2$, therefore

\be h_{ijkl} = \hp_i\hp_j\hp_k \hp_l\, \beta_1 + \hp_{(i}\hp_j\hp_{k}\p_{l)} \,\beta_2  \to  (I_{2s},I_{2s-1}) = (\Box^2\beta_1, - \Box\,\dot{\beta}_2)
 \label{beta12} \ee

\bea \lst &&= b_0 \, h_{ijkl}\Big[ \theta^{im}\theta^{jn}\theta^{kp}\theta^{lq} - \theta^{ij}\theta^{mn}\theta^{kp}\theta^{lq} + \frac{\theta^{ij}\theta^{mn}\theta^{kl}\theta^{pq}}8\Big]h_{mnpq} \label{i8a}\\ &&= b_0 \Big[ \beta_1 \nabla^8 \left(1-1+\frac 18 \right)\Box^4\beta_1 + \hp_{(i}\hp_j\hp_{k}\p_{l)} \,\beta_2  \left( \theta^{im}\theta^{jn}\theta^{kp}\theta^{lq} - \theta^{ij}\theta^{mn}\theta^{kp}\theta^{lq} \right)\hp_{(m}\hp_n\hp_{p}\p_{q)} \,\beta_2 \Big]\nn\\
&&= b_0 \Big[I_{2s} \frac{\nabla^8}8 I_{2s} + 2\, I_{2s-1} \, \nabla^8\, \Box \, I_{2s-1} \Big] \label{il8} \eea

\no where we have used $\p_i\p_m\Box\theta^{im} = \ns\p_0^2 $. Notice that no cross term $\beta_1 \times \beta_2$ appears due to the odd number of dual derivatives $\hp$ which leads to $\hp \cdot \p = 0$. Regarding the 7th order Chern-Simons term (\ref{cs4}) we have ,

\bea \lsm & &= \frac {c_0}2 h_{ijkl}\left( 2\, \theta^{jn}\theta^{kp}\theta^{lq} - \theta^{np}\theta^{jk}\theta^{lq}\right) \Box^3\dot{h}_{i(npq)}^* \label{i7a}\\
& &= -c_0\, \hp_j\hp_k\hp_l \beta_1 \left( 2\, \delta^{jn}\delta^{kp}\delta^{lq} - \delta^{np}\delta^{jk}\delta^{lq}\right)\ns \hp_n\hp_p\hp_q \dot{\beta}_2  = - c_0 \, I_{2s-1} \, \nabla^8 I_{2s} \label{il7} \eea

\no where we have used $E^{im}=\epsilon^{im}\p_0$ and

\be \dot{h}_{i(npq)}^* \equiv - \epsilon^{im}h_{mnpq} = - \p_i\hp_n\hp_p\hp_q \, \dot{\beta}_1 + \hp_i\hp_n\hp_p\hp_q \, \dot{\beta}_2 - \p_i\big[ \p_{(n}\hp_p \hp_{q)}\dot{\beta}_2\big] \quad . \label{hdual4} \ee

\no As in the spin-3 case, only the cross terms $\beta_1 \times \beta_2$ survive in (\ref{il7}) due to $\p \cdot \hp =0$.

Comparing (\ref{lst2})-(\ref{lsm2}) with (\ref{il8}) and (\ref{il7}) we see that the canonical structure of spin-2 and spin-4 cases basically coincide.  So the linearized NTMG (\ref{ntmg})  has its spin-4 counterpart, first suggested in \cite{yin}, with $c_0= - m\, b_0$ we have

\bea  \lsdt  &&=  b_0 \Biggl\{ h_{\mu\nu\alpha\beta}\Box^4 \Big[ \theta^{\mu\rho}\theta^{\nu\gamma}\theta^{\alpha\lambda}\theta^{\beta\sigma} - \theta^{\mu\nu}\theta^{\rho\gamma}\theta^{\alpha\lambda}\theta^{\beta\sigma} + \frac 18 \theta^{\mu\nu}\theta^{\rho\gamma}\theta^{\alpha\beta}\theta^{\lambda\sigma}\Big]h_{\rho\gamma\lambda\sigma} \nn\\
&& +m\,  h_{\mu\nu\alpha\beta}\square^{3}{E}^{\mu\gamma}\left(\theta^{\nu\rho}\theta^{\alpha\lambda}\theta^{\beta\sigma}-\frac{1}{2}\theta^{\nu\alpha}\theta^{\rho\lambda}\theta^{\beta\sigma}\right)h_{\gamma\rho\lambda\sigma} \Biggr\} \label{sd8} \\
&& =  2\, b_0 \, \left\lbrace \frac 14 I_{2m}\, \nabla^4\, I_{2m} +  \md\left\lbrack \nabla^4 (\Box - m^2 )\right\rbrack \md\right\rbrace \label{ntmg4} \eea

\no where $\lst$ is given in (\ref{i8a}) while  $\lsm$ appears in (\ref{i7a}), moreover  $I_{2m}=\tp/2+2m\, \md$. We can always change  variables $(I_{2m},I_{2s-1})=(\orho/2,\ogamma)$ and treat the two WDiff invariants as independent degrees of freedom. So we have just one massive mode in $\lsdt$ as shown in \cite{yin} in a fixed gauge.

 In order to find the spin-4 analogues of TMG and NMG we first need $\lsb$,

 \be \lsb = \int \, d^3x \, h_{\mu\nu\alpha\beta} D^{\mu\nu\alpha\beta}(h) \quad . \label{lsb4} \ee

 \no where $D^{\mu\nu\alpha\beta}(h)\sim \p^6 h $ is fully symmetric and satisfies

 \be C_{\mu\nu\alpha\beta} = E_{(\mu}^{\!\!\quad \rho}D_{\rho\nu\alpha\beta )}  \quad . \label{c4} \ee

 \no where $C_{\mu\nu\alpha\beta}$ is the spin-4 Cotton tensor in flat space  given in (\ref{cs4}). As in the spin-3 case we can alternatively start from a rather general Ansatz for $\lsb$ with all possible contractions and require $\odl \lsb =0$ where the $\odl$ transformations correspond to (\ref{gt5}) with all possible scalar restrictions on the gauge parameters,

 \be \p^{\alpha}\eta^{\mu\nu}\Lambda_{\mu\nu\alpha} \equiv \p\cdot\Lambda =0 = \p^{\mu}\p^{\nu}\p^{\alpha}\Lambda_{\mu\nu\alpha}=\p^{\mu}\p^{\nu}\psi_{\mu\nu} = \eta^{\mu\nu}\psi_{\mu\nu}  \quad . \label{r2} \ee

\no The general solution to the $\odl$ symmetry, or equivalently to (\ref{c4}),  is given by a five parameter family of Lagrangians

\bea  \lsb [a,b,c,d,e]\!\!\!\!\! &&= d_0\Big[ h_{\mu\nu\alpha\beta}\, \Box^3 h^{\mu\nu\alpha\beta} - h_{\mu\nu}\, \Box^3 h^{\mu\nu} + 4 \p_{\mu}h^{\mu\nu\alpha\beta}\, \Box^2 \p^{\lambda}h_{\lambda\nu\alpha\beta} - 2\, \p_{\mu}h^{\mu\nu} \, \Box^2 \p^{\lambda}h_{\lambda\nu} \nn\\
&&+ 2 \, h_{\mu\nu}\,\Box^2\p_{\alpha}\p_{\beta}h^{\mu\nu\alpha\beta} + 5\, \p_{\mu}\p_{\nu}h^{\mu\nu\alpha\beta}\, \Box \p^{\lambda}\p^{\sigma}h_{\lambda\sigma\alpha\beta} +4\, \p_{\mu}h^{\mu\nu}\, \Box \, \p^{\alpha}\p^{\beta} \p^{\lambda}h_{\lambda\nu\alpha\beta} \nn \\\
&& + 2\, \p^{\alpha}\p^{\beta} \p^{\lambda}h_{\lambda\nu\alpha\beta} \p_{\mu}\p_{\rho}\p_{\gamma}
h^{\nu\mu\rho\gamma}  + a\, h\, \Box^3 h + b\, h\, \Box^2 \p_{\mu}\p_{\nu}h^{\mu\nu}  + c\, \p_{\mu}\p_{\nu}h^{\mu\nu}\, \Box \, \p^{\lambda}\p^{\sigma}h_{\lambda\sigma} \nn\\ &&+ d\, h\, \Box \, \p_{\mu}\p_{\nu}\p_{\lambda}\p_{\rho} h^{\mu\nu\lambda\rho} + e\, \p_{\mu}\p_{\nu}h^{\mu\nu} \p_{\alpha}\p_{\beta}\p_{\lambda}\p_{\rho} h^{\alpha\beta\lambda\rho}\Big] \, , \label{l6}\eea

\no where $h\equiv \eta^{\mu\nu}h_{\mu\nu} \equiv \eta^{\mu\nu}\eta^{\alpha\beta} h_{\mu\nu\alpha\beta}$.

 Once again we look for a subset of solutions with an empty spectrum by requiring the maximal possible symmetry. First we have checked that there is no solution invariant under full WDiff (\ref{gt5}) constrained by only one of the restrictions (\ref{r2}). However, we have found at least five sets of two restrictions for which all
coefficients $(a,b,c,d,e)$ are fixed, they will be discussed elsewhere \cite{unp}. Here we only analyse the case:

\be \p \cdot \Lambda =0= \p^{\mu}\p^{\nu}\psi_{\mu\nu}  \to (a,b,c,d,e)=(\frac 3{25},-\frac 25,-1,\frac 15, 2) \quad . \label{5p} \ee

\no For convenience we define

\be {\cal L}_6^{(4)} \equiv \lsb \Big[\frac 3{25},-\frac 25,-1,\frac 15, 2\Big] \quad {\rm at} \quad d_0 = -m^2 b_0 \quad . \label{l64} \ee

\no At first sight (\ref{5p}) does not seem to be a perfect spin-4 analogue of (\ref{2p}). However, it turns out that if we start from a general Lagrangian of the form (\ref{l6}) but with all 13 coefficients arbitrary and require invariance under (\ref{gt5}) with the restrictions $(\Lambda_{\mu},\p^{\mu}\p^{\nu}\psi_{\mu\nu} )=(0,0)$ we would arrive exactly at (\ref{5p}). Likewise, in the spin-3 case, we have checked that if we start from a fourth order Lagrangian of the form (\ref{l4}) but with all 7 coefficients arbitrary and require symmetry under (\ref{gt4}) with the restrictions
$(\Lambda,\p \cdot \psi )=(0,0)$ we would end up precisely with (\ref{2p}). This means that instead of finding the higher spin analogues of the EH theory by searching for the solutions to (\ref{dtensor}) which have an empty spectrum, we can use instead a gauge symmetry principle just like the EH theory is completely fixed, up to trivial field redefinitions, by requiring Diff symmetry.

Henceforth we take (\ref{5p}) for granted. Note that the transformations (\ref{gt5}) restricted by  $\p^{\mu}\Lambda_{\mu}=0=\p^{\mu}\p^{\nu}\psi_{\mu\nu} $ still have a vector redundancy of the type discussed after (\ref{gt5}) but the vector must be transverse $\p^{\mu}\epsilon_{\mu}=0$. This means that we have in total $10+6-2-2=12$ independent gauge parameters\footnote{Alternatively, the four restrictions $\Lambda_{\mu}=0=\p^{\mu}\p^{\nu}\psi_{\mu\nu}$  also give 16-4=12 gauge parameters, no redundancy is left in this case.} in (\ref{gt5}) which leads to $15-12=3$ gauge invariants just like the previous $s=1,2,3$ cases. Solving (\ref{gt5}) for the 12 parameters and plugging back in (\ref{gt5}) we obtain three gauge invariants $\odl i_{2s}=0= \odl i_{2s-1}=\odl \oi_{2s-2}$. Besides the known invariants of 8th and 7th order given in (\ref{i8})  and (\ref{i7}) we have the 6th order invariant:

\bea \oi_{2s-2} &=& - \Box \, \left( \p_i\p_j\p_k\p_l h_{ijkl} - 2\, \ns \p_j\p_k h_{00jk} + \nabla^4 h_{0000} \right) + \Box\Big[8\p_i\p_j\hp_k\hp_l h_{ijkl} + 2 \nabla^2 \hp_j\hp_k h_{00jk} \Big] \nn\\
&&- \Box\, \hp_i\hp_j\hp_k\hp_l h_{ijkl}-10\ns\Big[ \p_i\p_j\hp_k\hp_l h_{ijkl} - 2 \, \p_j\hp_k\hp_l h_{0jkl} + \ns \, \hp_j\hp_k h_{00jk} \Big] \label{i64} \eea

\no In terms of helicity variables we have

\be \oI_{2s-2} = \frac{\oi_{2s-2}}{\nabla^6} = 20\nabla^2\dot{\psi}_2 - \frac{\Box}{\ns}\rho + 2\, \Box\phi_3 -2(\ddot{\phi}_1+4\ns\phi_1) - \ns\Box(\beta_1+\beta_5)-2\nabla^4\beta_3 -8\ns\ddot{\beta}_3 \nn\\ \label{I64} \ee

\no In order to write down $\lsb$ in terms of gauge invariants we suppose that  $\lsb=\sum_{K,L} I_K \hat{O}_{KL} \, I_L$
where the sum run over the 3 invariants (\ref{I8}),(\ref{I7}) and (\ref{I64}). If we first assume that the only non vanishing fields are ($\beta_1,\beta_2$) and then ($\beta_1,\beta_5$) , direct substitution in (\ref{l64})  with the constants given in (\ref{5p}) lead respectively to

\bea \lsb [\beta_1,\beta_2] &&= \frac{3}{25} \beta_1\, \nabla^8 \Box^3 \beta_1 - 2 \, \beta_2 \, \nabla^8 \Box^2 \ddot{\beta}_2  \quad , \label{E1} \\ \lsb [\beta_1,\beta_5] &&= \frac 15 \tilde{\beta}_1 \nabla^{10}\Box(\ns-2\Box)\beta_5 + \frac 3{25} \tilde{\beta}_1 \nabla^{8}\Box^3\tilde{\beta}_1 \quad . \label{E2} \eea

\no where $ \tilde{\beta}_1 = \beta_1 + \beta_5$. Moreover, if we keep only $(\gamma,\Gamma)$,  the only non vanishing components will be $h_{000j}$ and they are such that it is impossible to have a cross term $\gamma\times \Gamma$, consequently $\hat{O}_{78}=0=\hat{O}_{87}$.  From (\ref{E1}),(\ref{E2}) and $\hat{O}_{78}=0$ we obtain,

\be \lsb = d_0\Big[ 2\, I_{2s-1}\, \nabla^8 I_{2s-1} - \frac 15 I_{2s}\, \nabla^6 \oI_{2s-2} - \frac 2{25}  \oI_{2s-2}\, \nabla^4 \Box  \oI_{2s-2} \Big] \quad . \label{il64} \ee

\no Comparing (\ref{il8}),(\ref{il7}),(\ref{il64}) with the corresponding formulae of the spin-2 case  (\ref{lst2}),(\ref{lsm2}),(\ref{leh}) there is a perfect match  after a harmless redefinition $(I_{2s},\oI_{2s-2}) \to (2\, I_{2s},5\,\oI_{2s-2})$ in (\ref{il64}). However, we still have to worry whether we can treat $(I_{2s},I_{2s-1},\oI_{2s-2})$ given in (\ref{I8}),(\ref{I7}) and (\ref{I64}) as independent degrees of freedom such that the higher time derivatives can be properly  hidden via a change of variables in order to avoid ghosts. This is indeed the case as we show now.

First we redefine $\gamma$ such that $I_{2s-1} = \ogamma$. This is a trivial one field redefinition that does not affect neither
$I_{2s}$ nor $\oI_{2s-2}$.
Next we redefine $\rho $ such that $I_{2s}=\orho$ which introduces higher time derivatives  of $\psi_2$ in $\oI_{2s-2}$. They can be cancelled out after $\Gamma = \oGamma - 3 \, \Box \, \psi_2 + (3/2)\dot{\phi}_3 $ which allows us to get rid also of time derivatives of $\phi_3$  via $\psi_2 = \overline{\psi}_2 - \dot{\phi}_3/(10\, \ns) $. Finally, we redefine $\phi_3$ such that $\oI_{2s-2}=2\,\ns \overline{\phi}_3$. In summary, we have a fivefold change of variables $\overline{\Phi}_I=M_{IJ}\Phi_J + F_J $, where $\Phi_J=(\rho,\Gamma,\psi_2,\phi_3,\gamma)$ and $F_J$ do not depend upon $\Phi_I$, which leads to $(I_{2s},I_{2s-1},\oI_{2s-2})=(\orho,\ogamma,2\,\ns \overline{\phi}_3)$. One can check that $\det M =1$.


Consequently we can define the spin-4 analogues of the spin-2 linearized TMG and linearized NMG. In the first case if we choose $(c_0,d_0)=(-m\, b_0,-m^2b_0)$ we have

\bea \lsdm &&\equiv {\cal L}_{LTMG}^{(s=4)} =  -m\, b_0\, h_{\mu\nu\alpha\beta}\square^{3}{E}^{\mu\gamma}\left(\theta^{\nu\rho}\theta^{\alpha\lambda}\theta^{\beta\sigma}-\frac{1}{2}\theta^{\nu\alpha}\theta^{\rho\lambda}\theta^{\beta\sigma}\right)h_{\gamma\rho\lambda\sigma} + {\cal L}_6^{(4)}   \label{ltmg4} \\ &&= 2\, m^2\, b_0 \left\lbrack \tilde{\oI}_{2s-2} \, \nabla^6 \, \Box \, \tilde{\oI}_{2s-2} - \md \, \nabla^8 \md - \tilde{I}_{2s} \, \nabla^8 \left(\tilde{\oI}_{2s-2} + \frac{\md}m\right)\right\rbrack \quad . \label{iltmg4} \eea

\no where ${\cal L}_6^{(4)}$ is the 6th order spin-4 Lagrangian in (\ref{l64}). The 3 gauge invariants $(\tilde{I}_{2s},\md,\tilde{\oI}_{2s-2})=(2\, I_{2s},\md,5\, \oI_{2s-2})$ can be obtained from (\ref{I8}),(\ref{I7}) and (\ref{I64}). We can repeat the arguments given after (\ref{ltmg}) and prove that (\ref{ltmg4}) has only one massive propagating mode. It is invariant under the WDiff transformations (\ref{gt5}) constrained by $\p \cdot \Lambda = 0= \p^{\mu}\p^{\nu}\psi_{\mu\nu}$.

In the second case of the spin-4 linearized NMG model we add the Lagrangians (\ref{sd8}) and (\ref{ltmg4}) such that the higher spin Chern-Simons term cancel out and we have

\bea \ldt \!\!\! &&\equiv {\cal L}_{LNMG}^{(s=4)}  = b_0 \,  h_{\mu\nu\alpha\beta}\Box^4 \Big[ \theta^{\mu\rho}\theta^{\nu\gamma}\theta^{\alpha\lambda}\theta^{\beta\sigma} - \theta^{\mu\nu}\theta^{\rho\gamma}\theta^{\alpha\lambda}\theta^{\beta\sigma} + \frac 18 \theta^{\mu\nu}\theta^{\rho\gamma}\theta^{\alpha\beta}\theta^{\lambda\sigma}\Big]h_{\rho\gamma\lambda\sigma} + {\cal L}_6^{(4)} \nn\\  \label{nmg4} \\
&&  = 2 b_0 \left\lbrack \md \, \nabla^6 (\Box -m^2)  \md + m^2 \bt \, \nabla^4 (\Box -m^2) \, \bt + \oI_{2s}\, \nabla^8 \oI_{2s} \right\rbrack  \label{inmg4} \eea

\no where $ \oI_{2s} =\tilde{I}_{2s} -2\, m^2 \tilde{\oI}_{2s-2}  $. The spin-4 NMG theory is also invariant under (\ref{gt5}) restricted by the scalar conditions $\p \cdot \Lambda = 0= \p^{\mu}\p^{\nu}\psi_{\mu\nu}$.

\section{Gauge invariants and the Cotton and D-tensor}

In all cases $s=1,2,3,4$ worked out here we have been able to find two invariants under (\ref{gt1}), $(i_{2s},i_{2s-1})$, which play an instrumental role. The transformations (\ref{gt1}) can be rewritten without loss of generality as,

\be \delta h_{\mu_1  \cdots \mu_s} = \p_{(\mu_1}\oL_{ \mu_2  \cdots \mu_s)} + \eta_{(\mu_1\mu_2} \psi_{\mu_3  \cdots \mu_s )} \quad . \label{gt6} \ee

\no where $\eta^{\mu_2\mu_3}\oL_{ \mu_2  \cdots \mu_s}=0$. Since all three tensors in (\ref{gt6}) are fully symmetric, their number of components in 3D is given by

\bea N_{\delta h} &&= \frac{3.4.\cdots (3+s-1)}{s !} = \frac{(s+1)(s+2)}2 \nn\\ N_{\oL} &&= \frac{3.4.\cdots (3+s-2)}{(s-1) !} -\frac{3.4.\cdots (3+s-4)}{(s-3) !} = 2\, s-1 \nn\\
N_{\psi} &&=  \frac{3.4.\cdots (3+s-3)}{(s-2) !} = \frac{s(s-1)}2 \nn \eea

\no Therefore we always have only two invariants under (\ref{gt6}) for arbitrary integer spin-s,

\be N_I = \frac{(s+1)(s+2)}2 - \Big[ 2\, s-1+ \frac{s(s-1)}2 \Big] = 2 \quad . \label{ni} \ee

\no On the other hand, the Cotton tensor $C_{\mu_1 \cdots \mu_s}$ is also fully symmetric, transverse and traceless, see (\ref{fpc}). Therefore, the same counting applies to $C_{\mu_1 \cdots \mu_s}$ which must have only two independent components invariant under (\ref{gt6}). This rises the question about the connection between the invariants and the Cotton tensor. For all cases presented here we have found $\lsm \sim 2\, c_0 \, i_{2s-1} \, \nabla^{-2\, s} \, i_{2s}= c_0 h_{\mu_1  \cdots \mu_s} \, C^{\mu_1\cdots \mu_s } $. Since $h_{00\cdots 0} $ and $\hp_j h_{00\cdots j}$ appear linearly  in $i_{2s}$ and in $i_{2s-1}$ respectively, from the functional derivatives $ \hp_j \delta S_{2s-1}/\delta  h_{00\cdots j}$ and $\delta S_{2s-1}/\delta h_{00\cdots 0} $ we learn that $(i_{2s},i_{2s-1})\sim (\hp_j C_{00\cdots j},C_{0\cdots 0}) $. We have confirmed in all cases $s=1,2,3,4$ that  this is indeed the case. Thus, we do not need to solve for the gauge parameters in (\ref{gt6}) in terms of $\delta h_{\mu_1 \cdots \mu_s}$  in order to obtain  $(i_{2s},i_{2s-1})$.

Whenever we consider the Lagrangian $\lsb$ the number of symmetries is decreased by one unit  and consequently we need one more gauge invariant which corresponds to $\mdb $ (odd spin) or to $\bt$ (even spin).  In the even spin cases $s=2,4$ we see that $I_{2s}$ appears linearly  in $\lsb$, see (\ref{leh}) and (\ref{il64}), since $\dot{\Gamma}$ is only present in
$I_{2s}$, see (\ref{i2bi3i4})  and (\ref{ltmg4}), it turns out that $\p_j D_{00\cdots j} \sim \dot{\oI}_{2s-2} $, so we have a short cut to obtain also $\oI_{2s-2}$. In the odd spin case we have not been able to find any short cut for  $\mdb$ which might avoid long  calculations involving the elimination of the gauge
parameters in the gauge transformations (\ref{gt1}).

\section{Conclusion}

Here we have suggested spin-4 analogues of the linearized TMG and linearized NMG, see (\ref{ltmg4}) and (\ref{nmg4}) respectively. Although those models are of 7th and 8th order in derivatives respectively, we have shown that they  are ghost free and moreover they have exactly the same canonical structure of their spin-2 counterpart when written in terms of appropriate gauge invariants, see (\ref{iltmg4}), (\ref{inmg4}) and compare with (\ref{ltmg}) and (\ref{lnmg}). The canonical structure of the linearized spin-4 NTMG (\ref{sd8}), suggested in \cite{yin}, also coincides with its spin-2 counterpart, compare (\ref{ntmg}) with (\ref{ntmg4}). We have also shown that the spin-3 linearized TMG, NTMG and NMG have the same canonical structure of the spin-1 first order self-dual model of \cite{tpn}, Maxwell Chern-Simons and Maxwell-Proca models respectively.

An important ingredient for higher spin linearized TMG and NMG is the D-tensor in the action $\lsb$. It is the tensor whose symmetrized curl is the Cotton tensor (\ref{dtensor}). This condition guarantees that the self-dual model of order 2s-1 built by combining $\lsm$ and $\lsb$, see (\ref{lsdm}), contains particles with helicity $s\vert m\vert /m$. However, there might be further particles including ghosts. In general, the condition (\ref{dtensor}) leads to a multi parametric family of Lagrangians $\lsb$. In the rank-2 case $\lsb$ becomes  the two parameter family of TDiff models \cite{blas} which in $D=2+1$  contains only a scalar field in the spectrum which might have a  wrong overall sign depending on the parameters of the model. Consequently the third order self-dual model defined in (\ref{lsdm}) might contain a scalar ghost besides the helicity $2\vert m \vert / m$ particle. In order to avoid such extra modes we have learned from previous works \cite{djt,dj,sd4,sd5,ghosh} that $\lsb$ must have an empty spectrum. In the spin-2 case this leads to only two possibilities. Namely, either TDiff is extended to Diff (Einstein-Hilbert) or to WTDiff (linearized unimodular gravity). The corresponding self-dual models become the TMG of \cite{djt} and the unimodular TMG of \cite{ghosh} respectively.

For higher spins $s \ge 3$ we are still investigating \cite{unp} possible candidates for $\lsb $ satisfying (\ref{dtensor}) and without particle content. There is however, at least one natural higher spin version of the linearized Einstein-Hilbert (LEH) theory in $D=2+1$ in the $s=3$ and $s=4$ cases. In the $s=3$ case it is the fourth order Lagrangian given in (\ref{l43}), see \cite{sd5}. It is invariant under (\ref{gt4}) restricted by $\Lambda = 0 = \p^{\mu} \cdot \psi_{\mu}$. The $s=4$ case corresponds to the 6th order Lagrangian in (\ref{l64}) which is invariant under (\ref{gt5}) with the restrictions $\p_{\mu}\Lambda^{\mu} =0=\p^{\mu}\p^{\nu}\psi_{\mu\nu}$. Just like the LEH theory in $D=2+1$, we have shown here, in a gauge invariant way, that both (\ref{l43}) and (\ref{l64}) have no particle content.

The Lagrangians ${\cal L}_{LEH}$, ${\cal L}_4^{(3)}$ and ${\cal L}_6^{(4)}$  share another interesting feature. The LEH theory is uniquely determined by its order in derivatives and invariance under linearized diffeomorphisms. Likewise (\ref{l43}) and (\ref{l64})  are uniquely determined by their order in derivatives and invariance under restricted $\Lambda$ and Weyl transformations. Namely, if we start with a Lagrangian of the form (\ref{l4}) but  with all 7 terms with arbitrary coefficients and require invariance under (\ref{gt4}) restricted by  $\Lambda = 0 = \p^{\mu} \psi_{\mu}$ we end up with ${\cal L}_4^{(3)}$. Similarly, beginning with a Lagrangian  of the form (\ref{l6}) with 13 arbitrary constants and demanding invariance under (\ref{gt5}) restricted by $\Lambda_{\mu} =0= \p^{\mu}\p^{\nu}\psi_{\mu\nu}$ we arrive precisely at ${\cal L}_6^{(4)}$. Notice that there is no need of requiring (\ref{dtensor}). From this point of view $\lsb $ is on the same footing of $\lst$ and  $\lsm$, they are all completely determined by their order in derivatives and a local symmetry, namely a restricted conformal higher spin symmetry. Moreover, $({\cal L}_{LEH},{\cal L}_4^{(3)},{\cal L}_6^{(4)})$ all have one less symmetry than WDiff (\ref{gt1}). It is tempting to generalize the above symmetry restrictions for arbitrary integer spins in order to infer an arbitrary spin-s canonical structure for all singlets of order $2s$ and $2s-1$ and doublets of order $2s$.

We point out that the method we have used here for investigating
the particle content of higher derivative theories dispenses the use of
gauge conditions which clarifies the underlying canonical structure.
Furthermore, it holds off-shell which is specially useful for the doublet models $\ldt$ with both helicities $\pm s$  where the relative sign of the two massive modes is crucial for absence of ghosts. We mention that the extension of the gauge invariant formulation to curved backgrounds is very promising, see \cite{jaccard}. Regarding the generalization of our self-dual models of order $2s-1$ to curved backgrounds see the recent work \cite{hkp}.

Instead of stepping down the ladder of higher spin self-dual models mentioned in the introduction as we have done here going from the eighth order spin-4 self-dual model to the seventh order one, one might try to go upstairs starting from the first order spin-4 self-dual model of \cite{aragone}. However, a recent attempt, see \cite{elias4}, along such direction gets stuck apparently at the fourth order model which is the same order at which the spin-3 case stops \cite{nge3}. One might try\footnote{We thank an anonymous referee for an inspring question about that point.} to go upstairs by either using the master action or the Noether gauge embedding procedure starting from the first order action of \cite{tv} for arbitrary spin self-dual models.

Finally, that the most general D-tensor solution to (\ref{dtensor}) in the spin-2 case leads to TDiff (transverse diffeomorphisms) theories. It is known that TDiff is the minimal symmetry for massless spin-2 particles in $D=3+1$ which can be related to massive particles of helicity $\pm 2$ in $D=2+1$ via dimensional reduction. Since for $s=3$ and $s=4$ the D-tensor definition
(\ref{dtensor}) can be traded into a symmetry principle under general transformations (\ref{gt1})  restricted by scalar constraints, see (\ref{r1}) and (\ref{r2}),  it may be worth to investigate the role of those symmetries in $D=3+1$.

\section{Acknowledgements}

The work of D.D. is partially supported by CNPq  (grant 306380/2017-0). A.L.R.dos S. has been supported by a CNPq-PDJ
(grant 160784/2019-0).

\end{document}